\documentclass[11pt]{iopart} 

\usepackage{lipsum}
\usepackage{iopams}
\usepackage{amsmath,amsfonts,amssymb, mhchem}
\usepackage{graphicx}
\usepackage{siunitx}
\usepackage{float}

\usepackage{lineno}

\usepackage{filecontents}

\usepackage{hyperref}
\hypersetup{hidelinks}
\usepackage{xr}
\usepackage[font=footnotesize,labelfont=bf]{caption}
\usepackage{wrapfig}
\usepackage[outercaption]{sidecap}
\makeatletter
\def\SC@figure@vpos{m}
\makeatother
\usepackage{tabularx}
\usepackage{color,soul}
\usepackage{bm}
\usepackage{titletoc}
\usepackage{caption}
\renewcommand{\figurename}{Fig.}
\usepackage{ulem}
\usepackage{matlab-prettifier}
\usepackage[backend=bibtex, style=nature]{biblatex}
\renewbibmacro{begentry}{\midsentence}
\AtBeginBibliography{\small}

\usepackage{etoolbox}
\usepackage{setspace}
\newcounter{suppfigure}
\newcommand{\listofsuppfigures}{\section*{Supplementary Figures}\@starttoc{lof}}

\DeclareCiteCommand{\citen}
  {}
  {\printfield{labelnumber}}
  {\multicitedelim}
  {}

\begin{document}
\singlespacing
\raggedbottom
\clearpage

\title[]{Bistable Organic Electrochemical Transistors: Enthalpy vs. Entropy}

\author{Lukas M. Bongartz$^{1*}$, Richard Kantelberg$^{1}$, Tommy Meier$^{1}$, Raik Hoffmann$^{2}$, Christian Matthus$^{3}$, Anton Weissbach$^{1}$, Matteo Cucchi$^{1}$, Hans Kleemann$^{1}$, Karl Leo$^{1}$}

\address{$^{1}$IAPP Dresden, Institute for Applied Physics, Technische Universit\"at Dresden, N\"othnitzer Str. 61, 01187 Dresden, Germany}
\address{$^{2}$Fraunhofer Institute for Photonic Microsystems IPMS, Center Nanoelectronic Technologies, An der Bartlake 5, 01099 Dresden, Germany}
\address{$^{3}$Chair of Circuit Design and Network Theory (CCN), Faculty of Electrical and Computer Engineering, Technische Universit\"at Dresden, Helmholtzstr. 18, 01069 Dresden, Germany}

\begin{abstract}
    Organic electrochemical transistors (OECTs) underpin a range of emerging technologies, from bioelectronics to neuromorphic computing, owing to their unique coupling of electronic and ionic charge carriers. In this context, various OECT systems exhibit significant hysteresis in their transfer curve, which is frequently leveraged to achieve non-volatility. Meanwhile, a general understanding of its physical origin is missing. Here, we introduce a thermodynamic framework that readily explains the emergence of bistable OECT operation via the interplay of enthalpy and entropy. We validate this model through temperature-resolved characterizations, material manipulation, and thermal imaging. Further, we reveal deviations from Boltzmann statistics for the subthreshold swing and reinterpret existing literature. Capitalizing on these findings, we finally demonstrate a single-OECT Schmitt trigger, thus compacting a multi-component circuit into a single device. These insights provide a fundamental advance for OECT physics and its application in non-conventional computing, where symmetry-breaking phenomena are pivotal to unlock new paradigms of information processing.
\end{abstract}
\section*{Introduction}
\noindent Organic electrochemical transistors (OECTs) are at the foundation of numerous emerging technologies, ranging from bioelectronic implants\autocite{someya2016rise, park2018self} to analog neuromorphic computing\autocite{van2017non, van2018organic, gkoupidenis2015neuromorphic, cucchi2023liquido}, and have recently also expanded into the realm of complementary circuitry\autocite{huang2023vertical}. This track record is tied to a unique switching mechanism, based on the coupling of ionic and electronic charge carriers in an organic mixed ionic-electronic conductor (OMIEC)\autocite{rivnay2018organic, paulsen2020organic, gkoupidenis2023organic}. Connected with two electrodes, source and drain, polarons can be driven along the channel by applying a corresponding voltage $V_\mathrm{DS}$, giving rise to a drain current $I_\mathrm{D}$. By immersing the system in an electrolyte along with a gate electrode, a second voltage $V_\mathrm{GS}$ enables control over the ion flow between electrolyte and channel, and as such, the doping level of the latter (Fig.\,\ref{fig:1}d). The benchmark channel material of OECTs is the polymer blend poly(3,4-ethylenedioxythiophene) polystyrene sulfonate (PEDOT:PSS), where the transistor switching can be described by 
\begin{equation}
\label{eq:pedot reaction}
      \ce{
PEDOT{:}PSS + C^+ + A^- <=> PEDOT^{+}{:}PSS + C^+ + A^- + e^-,
}  
\end{equation}
with \ce{PEDOT{:}PSS} and \ce{PEDOT^{+}{:}PSS} as the initial and doped state of the OMIEC. \ce{C^+} and \ce{A^-} are the electrolyte cat- and anions and most commonly are given by liquid electrolytes such as aqueous NaCl solution. For a long time, OECTs have been understood by concepts borrowed from field-effect transistor (FET) theories, in particular based on the foundational work by Bernards and Malliaras\autocite{bernards2007steady}. At the same time, however, charge formation and transport in FETs differs fundamentally from those in OECTs. While in the former, charges are induced at the semiconductor interface and form a quasi-2D layer, doping in OECTs happens throughout the entire channel, where the formation of electrical double layers at the polymer strands results from an electrochemical process\autocite{friedlein2018device}. Cucchi et al. have set out to resolve this mismatch by describing OECTs in terms of thermodynamics, thereby addressing the long-standing question of the non-linear voltage dependence of the capacitance{\autocite{cucchi2022thermodynamics}}. In particular, this work describes the process of Eq.\,\ref{eq:pedot reaction} as being driven by entropy, where the gate voltage $V_\mathrm{GS}$ determines the electrochemical potential and as such shifts the equilibrium charge carrier concentration in PEDOT. However, these considerations are limited to the unidirectional operation of the OECT and pertain to the edge case where minimal or no interactions are assumed.

In this work, we move beyond this limitation. We consider the interactions involved in the doping cycle in terms of enthalpy, which allows us to reveal the peculiar situation of a bistable switching behavior. We do so by moving away from aqueous electrolytes to a solid-state system, in order to expose otherwise screened interactions and their effect on the Gibbs free energy. We establish a mathematical condition for the degree of bistability and underpin our reasoning with three sets of experimental evidence. Among others, we show the exceptional scenario in which the subthreshold swing behaves contrary to what would be expected from Boltzmann's law. We re-interpret results from the literature in the context of this framework, which further supports our model and allows for an alternative view on material properties for neuromorphic applications. Finally, our insights culminate in demonstrating how the bistable switching behavior can be leveraged to realize the functionality of a Schmitt trigger -- a multi-component circuit -- through a single device, laying the foundation for more complex circuitry in neuromorphic and asynchronous computing.

\section*{Results}
\subsection*{Theoretical Framework}

\begin{figure}[t!]
    \centering
    \includegraphics[width=\linewidth]{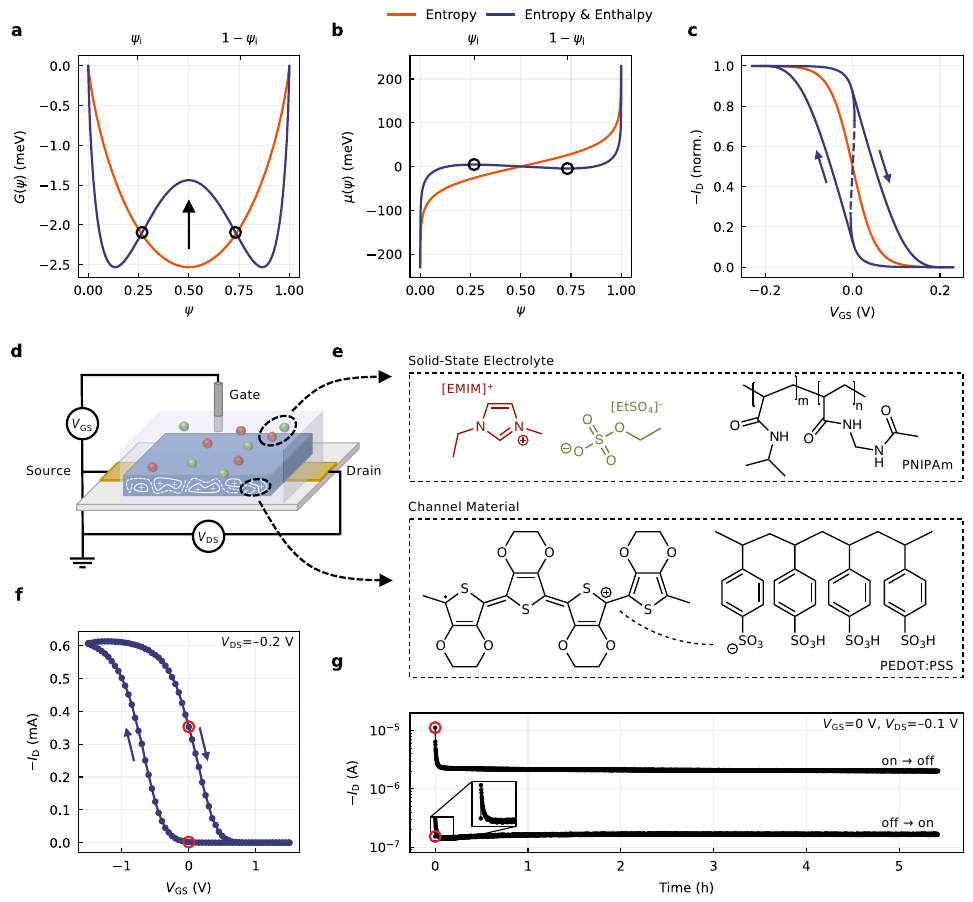} 
    \caption{\textbf{Bistability emerges from the interplay of enthalpy and entropy.} (\textbf{a}-\textbf{c}) The Gibbs free energy function $G(\psi)$ governs the chemical potential $\mu(\psi)$, which itself determines the transfer curve of an OECT. In case of dominating entropy, the system is monostable along the $\psi$-axis, with a monotonic chemical potential and an accordingly shaped transfer curve (orange). For rising enthalpy, the system gets bistable, causing a non-monotonic chemical potential, which results in a bistable switching behavior (blue). (\textbf{d}) Generalized setup of an OECT. Devices in this work typically employed a side-gate architecture (Fig.\,{\ref{fig:S_OECT_Figures}}). (\textbf{e}) Chemical structures of the OECT solid-state electrolyte and channel material used to study the bistability, with (\textbf{f}) a representative, experimental transfer curve ($V_\mathrm{DS}=-0.2\,\si{\volt}$). (\textbf{g}) Two distinct drain current levels are found when operating the OECT and holding a 0\,\si{\volt} gate bias, approaching either from the on- or the off-state. 
    \label{fig:1}}
     \vspace*{-\baselineskip}
\end{figure}

The Gibbs free energy is defined as the difference between enthalpy and entropy scaled by temperature, according to
\begin{equation}
    G = H - TS,
\end{equation}
where $H$ is enthalpy, $T$ is temperature, and $S$ is entropy. Akin to Cucchi et al.\autocite{cucchi2022thermodynamics}, we consider the Gibbs free energy as a function of an intensive state variable, here being the doping parameter $\psi$. Let the channel be composed of doping subunits, where one subunit is the smallest entity that satisfies Eq.\,\ref{eq:pedot reaction}, then $\psi$ describes the ratio of doped units ($N_\mathrm{doped}$) to the total number available ($N_\mathrm{tot}$). In other words, it serves as a statistical measure and reflects the probability distribution of doped units across all microstates:
\begin{equation}
    \psi = \frac{N_\mathrm{doped}}{N_\mathrm{doped}+ N_\mathrm{undoped}}= \frac{N_\mathrm{doped}}{N_\mathrm{tot}}.
    \label{eq:def_phi}
\end{equation}
The OECT switching can be understood as controlling this very ratio. For such a process, the Gibbs free energy (per unit) follows as 
\begin{equation}
     G(\psi) = H^0(\psi) + H_\mathrm{tr}(\psi) - TS_\mathrm{tr}(\psi),
     \label{eq:Gibbs_Intro}
\end{equation}
where $H^0(\psi)$ is the standard enthalpy. $H_\mathrm{tr}(\psi)$ and $S_\mathrm{tr}(\psi)$ are the enthalpy and entropy associated with the state transitions of Eq.\,{\ref{eq:pedot reaction}}, defined as
\begin{align}
H_\mathrm{tr}(\psi) &= \frac{1}{2}\left(h_{dd}\psi^2 + h_{uu}(1-\psi)^2 + 2h_{du} \psi(1-\psi)\right)  \quad \text{and} \label{eq:H_mix} \\
S_\mathrm{tr}(\psi) &= -k_\mathrm{B} \left(\psi\ln(\psi) + (1-\psi)\ln(1-\psi)\right).\label{eq:S_mix}
\end{align}
Here, $h_{dd}$, $h_{uu}$, and $h_{du}$ denote the intra- and interspecies interaction strength of doped and undoped sites and $k_\mathrm{B}$ is the Boltzmann constant (see \hyperref[Note_S:Theory]{Supplementary Note 1} for a comprehensive derivation). Both, the doping parameter $\psi$ and the Gibbs free energy $G(\psi)$ can be translated into device-level quantities. $\psi$ is proportional to the polaron density and therefore electrical conductivity (drain current), while the chemical potential $\mu(\psi)$ links $G(\psi)$ to the gate-source voltage over the electrochemical potential $\bar{\mu}$:
\begin{equation}
\label{eq:Chem_Potential}
 \left(\frac{\partial G(\psi)}{\partial \psi}\right)_{p,T} = \mu(\psi) \leftrightarrow \bar{\mu}  = \mu(\psi) + fe(V_\mathrm{GS}-V_\mathrm{Ch}),
\end{equation}
with $e$ as the elementary charge and $V_\mathrm{Ch}$ as the channel potential. $f$ is a fudge factor related to the effective modulation of the chemical potential through the gate-source voltage (doping efficiency) and underlies the translation from the theoretical energy scales (Fig.\,{\ref{fig:1}}b) to the experimentally found voltages (Fig.\,{\ref{fig:1}}f). In consequence of Eq.\,{\ref{eq:Chem_Potential}}, one can understand the switching characteristics of OECTs as a natural consequence of their underlying Gibbs free energy profile.

\nocite{bistability_simulation}
For a system of negligible interaction (enthalpy), entropy is the underlying driving force. As depicted in Fig.\,\ref{fig:1}a (orange), such systems possess a single minimum in $G(\psi)$, i.e., a single equilibrium state at $\psi_0$. It follows a monotonic chemical potential (Fig.\,\ref{fig:1}b), which translates to the transfer curve and determines the characteristic saturation in the on- and off-states of OECTs (Fig.\,\ref{fig:1}c)\autocite{cucchi2022thermodynamics}. This situation changes distinctively, when enthalpic contributions are taken into account. As derived in \hyperref[Note_S:Theory]{Supplementary Note 1}, the single equilibrium state in $G(\psi)$ bifurcates for

\begin{equation}
\lambda = \frac{(h_{dd}+h_{uu}-2h_{du})}{k_\mathrm{B}T} \cdot \psi(\psi-1)  \geq 1 \quad \text{with} \quad \psi\in[\psi_i, 1-\psi_i],
\label{eq:Condition_Bistability}
\end{equation}
where $G(\psi)$ has a negative curvature. Doping concentrations in this $\psi$-range are unstable and break into two coexisting equilibrium states along the doping axis, where $\psi_i$ and $1-\psi_i$ define the local extrema of the non-monotonic chemical potential. This thermodynamic instability goes along with a dynamic instability (\hyperref[Note_S:Dynamics_Instability]{Supplementary Note 2}), which inevitably suggests a bistable switching behavior (Fig.\,\ref{fig:1}a-c, blue). The quantity $\lambda$ sets the enthalpic and entropic contributions into relation and thus serves as a bifurcation parameter, reflecting the degree of bistability present in the system (Fig.\,\ref{fig:S_Bifurcation}). We provide an interactive simulation tool under Ref.\,\citen{bistability_simulation} to illustrate these relationships.

\subsection*{Theoretical Implications}

The bistability appears as the result of dominating enthalpy over entropy, which itself originates from the interactions underlying Eq.\,\ref{eq:H_mix}. It is reasonable to assume that their impact is most observable in a non-aqueous system, where dielectric shielding is minimized. At the same time, the ionic species themselves would benefit from a particularly strong dipole moment in order to penetrate and dope the OMIEC in the absence of water. For this purpose, we turn to the previously reported OECT system (Fig.\,{\ref{fig:1}e}, {\ref{fig:S_OECT_Figures}}) of PEDOT:PSS and 1-ethyl-3-methylimidazolium ethyl sulfate (\ce{[EMIM][EtSO_4]}) in a matrix of poly(N-isopropylacrylamide) (\ce{PNIPAm})\autocite{weissbach2022photopatternable}, offering multiple advantages: First, the solid-state system can be channeled into an inert gas atmosphere without degradation, providing a controlled, water-free environment. Second, \ce{[EMIM][EtSO_4]} offers among the highest dielectric permittivity of commercially available ionic liquids\autocite{weingaertner2014static}, and third, is known to provide an exceptionally low-lying off-state in OECTs\autocite{weissbach2022photopatternable}, allowing for a particularly large modulation of the charge carrier concentration. That is, a large range over which the $\psi$-axis can be probed. 

As demonstrated in previous reports\autocite{weissbach2022photopatternable, shameem2022hysteresis}, this system does in fact show a significant hysteresis in its transfer curve (Fig.\,\ref{fig:1}f), which has been shown to remain even for scan rates below $10^{-4}\,\si{\volt\per\second}$ (Ref.\,{\citen{weissbach2022photopatternable}}) and is similarly present with a non-capacitive Ag/AgCl-gate electrode (Fig.\,{\ref{fig:S_Transfer_Slow}}, {\ref{fig:S_AgAgClGate}}). Both of these findings suggest a reason beyond ion kinetics or capacitive effects\autocite{kaphle2018organic, ohayon2023guide, bisquert2023hysteresis} and motivate to study the presence of coexisting equilibrium states. To this end, we operate an OECT multiple times before holding a 0\,V gate bias, approaching either from the on- or the off-state (Fig.\,\ref{fig:S_Normal_longterm}). As Fig.\,\ref{fig:1}g shows, after a short stabilizing period ($\sim\si{\milli\second}$), two distinct drain current levels are present over the course of hours, differing more than one order of magnitude. Crucially, the gate current stabilizes at the same level ($\sim\si{\nano\ampere}$) for both conductivity states (Fig.\,{\ref{fig:S_Normal_longterm_GateCurrent})}. This result clearly indicates the presence of two separate doping states of the channel, present at the same gate bias but originating from different initial states. 
\begin{figure}[t!]
    \centering
    \includegraphics[width=\linewidth]{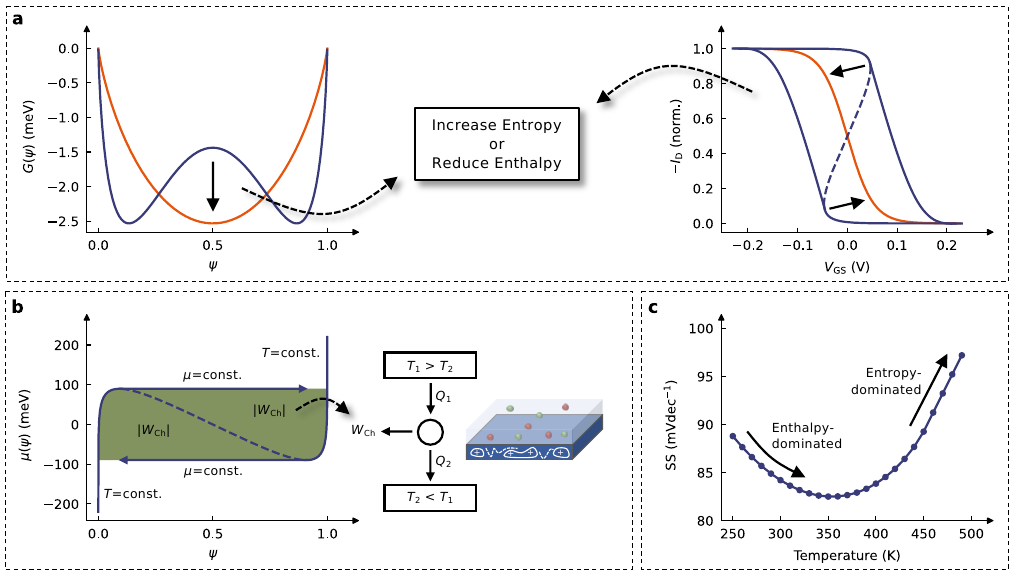} 
    \caption{\textbf{Theoretical implications of a bistable Gibbs free energy function.} (\textbf{a}) The bistability is expected to decrease for rising entropy ($TS_\mathrm{tr}$) and falling enthalpy ($H_\mathrm{tr}$), which should reflect in the transfer curve by an abating hysteresis. (\textbf{b}) The course of $\mu(\psi)$ can be considered an idealized cycle process, assembled of processes with $dT=0$ and $d\mu=0$. Following a Maxwell construction, the integrals correspond to the chemical work performed ($|W_\mathrm{Ch}|$), from which an in- and outflow of heat $Q$ is to be expected. (\textbf{c}) The bistability suggests a non-monotonic progression of the subthreshold swing ($SS$) with temperature, where the conventionally expected positive slope only occurs once entropy dominates over enthalpy and the chemical potential is sufficiently monotonic (Eq.\,\ref{eq:OECTSS}). \label{fig:2}}
     \vspace*{-\baselineskip}
\end{figure} 

We validate the hypothesis of a bistable Gibbs free energy function based on the theoretical consequences that this situation would entail for the ensemble. To this end, three hypotheses are derived (Fig.\,\ref{fig:2}). 

To begin with, given that the bistability is attributed to the balance between entropy and enthalpy (Eq.\,\ref{eq:Condition_Bistability}), we conclude that tweaking these quantities should allow for a targeted manipulation of the hysteresis in the transfer curve. Both an increasing entropy ($TS_\mathrm{tr}$) and a decreasing enthalpy ($H_\mathrm{tr}$) should cause the hysteresis to abate, as $\lambda$ decreases (Fig.\,\ref{fig:2}a). 

Second, as shown in \hyperref[Note_S:Maxwell]{Supplementary Note 3}, it is possible to think of the course of $\mu(\psi)$ as an idealized cycle process built up from processes of $dT=0$ and $d\mu=0$. With a non-monotonic chemical potential $\mu(\psi)$, and using the concept of Maxwell constructions, the enclosed area (per switching operation) then corresponds to the chemical work performed ($|W_\mathrm{Ch}|$), which in turn should be reflected in an in- and outflow of heat $Q$ (Fig.\,\ref{fig:2}b). 

Finally, these considerations can be extended to the subthreshold behavior. With the total chemical potential remaining constant during the transition between the on- and off-state, it follows that in this region ($\psi\in[\psi_i, 1-\psi_i]$), the non-monotonic chemical potential counterweights the electrostatic potential from the gate voltage (Eq.\,\ref{eq:Chem_Potential}). As the non-monotonicity decreases with temperature, this balancing effect is expected to decrease similarly, until it vanishes completely once entropy balances out. As shown in \hyperref[Note_S:SubthresholdSwing]{Supplementary Note 4}, we infer that for a bistable system, such a behavior should reflect in the subthreshold swing ($SS$) being subject to two effects: the counterbalance of the non-monotonic chemical potential as well as the classical influence of diffusion through thermal energy. Since the former is expected to decrease with rising temperature, while the latter scales with $k_\mathrm{B}T$, a non-monotonic progression of the subthreshold swing with temperature follows. This relationship can be expressed through
\begin{equation}
\label{eq:OECTSS}
    \mathrm{SS}(T)  \approx \frac{\ln(10)}{e} \left( \left(\left. \frac{\partial G(\psi, T)}{\partial \psi} \right|_{\psi =\psi_i}\right)_{p,T} +k_\mathrm{B}T \right),
\end{equation}
where the chemical potential is examined at the lower inflection point of $G(\psi)$, i.e., at its local maximum, based on the depletion mode operation of PEDOT:PSS. This equation is modelled in Fig.\,\ref{fig:2}c, showing that the conventionally expected increase of the subthreshold swing with temperature only occurs once the bistability is sufficiently suppressed by entropy. By the same token, we suggest a decreased subthreshold swing for a system of reduced enthalpy, which similarly lessens the non-monotonic nature of the chemical potential profile.

\begin{figure}[t!]
    \centering
    \includegraphics[width=\linewidth]{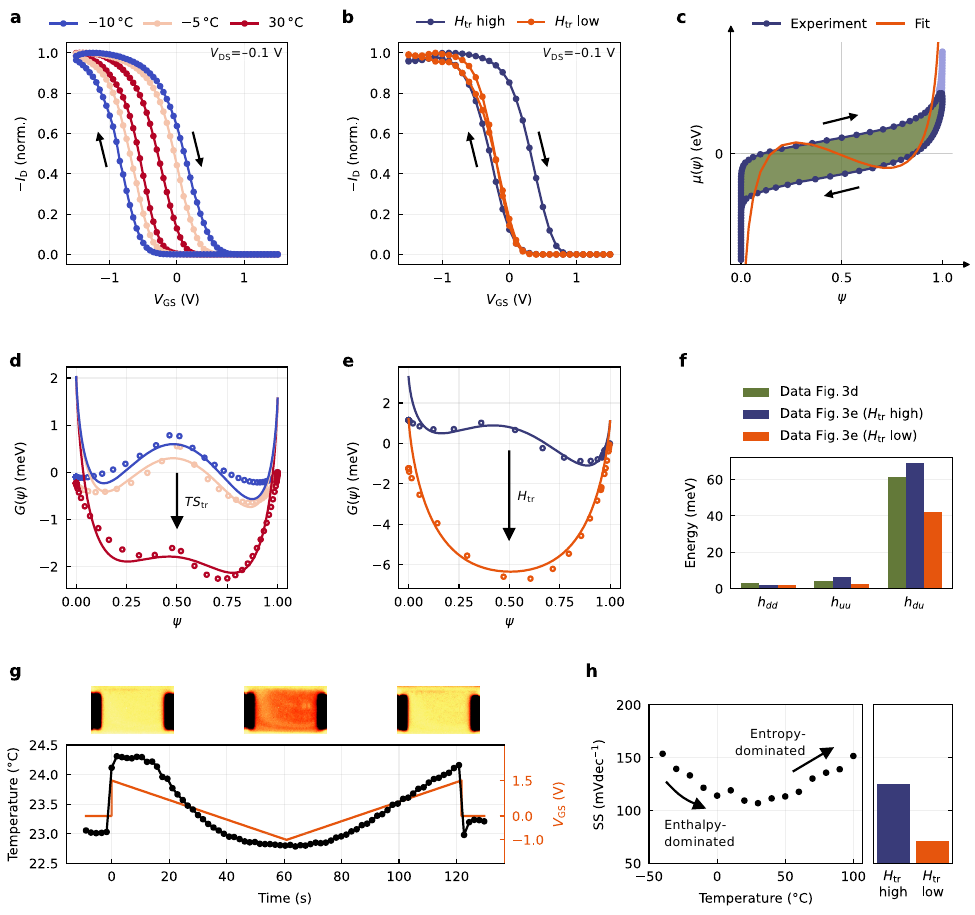} 
    \caption{\textbf{Experimental validation.} (\textbf{a,b}) Bistability decreases with (\textbf{a}) rising entropy and (\textbf{b}) falling enthalpy ($H_\mathrm{tr}$ high: untreated device, $H_\mathrm{tr}$ low: device with added KCl), in line with the expectations from Fig.\,\ref{fig:2}a ($V_\mathrm{DS}=-0.1\,\si{\volt}$). (\textbf{c}) Chemical potential profiles are normalized with symmetry around $(\psi, \mu(\psi))=(0.5, 0\,\si{\electronvolt})$ and partial integrals of equal size. Analytic continuation was performed to account for the asymmetry of the on- and off-state (light blue data points). The fit corresponds to the chemical potential derived from the fitted Gibbs free energy. (\textbf{d},\textbf{e}) Reconstructing and fitting $G(\psi)$ reveals decreasing local maxima for both experiments of (\textbf{a}) and (\textbf{b}), with the extracted interaction parameters summarized in (\textbf{f}). (\textbf{g}) Thermal imaging reveals the in- and outflow of heat during device operation, affirming the conclusion from Fig.\,\ref{fig:2}b. (\textbf{h}) The non-monotonic progression of the subthreshold swing with temperature confirms the hypothesis of Fig.\,\ref{fig:2}c. Further, the subthreshold swing decreases for reduced enthalpy as suspected.\label{fig:3}}
     \vspace*{-\baselineskip}
\end{figure}

\subsection*{Experimental Validation}
We verify these three hypotheses experimentally. To start with, we study the impact of entropy via temperature-dependent transfer measurements in quasi-steady state\autocite{bongartz2022temperature}. As shown in Fig.\,\ref{fig:3}a, these clearly confirm a decreasing bistability for rising temperature, which is also particularly apparent in logarithmic scale (Fig.\,\ref{fig:S_KCl_log}a). To validate the same for lowered enthalpy, we modify the electrolyte system by including KCl as a hygroscopic additive\autocite{jing2017hygroscopic}, thus shielding the interactions underlying $H_\mathrm{tr}$. As evidenced in Fig.\,\ref{fig:3}b, this attempt likewise confirms our hypothesis. Strikingly, the modification does not alter the on- or off-state of the OECT (Fig.\,\ref{fig:S_KCl_log}b) and solely causes the hysteresis curve to close. Using this system, we replicate the experiment of Fig.\,{\ref{fig:1}}g, confirming that the lapsed hysteresis translates to a diminished state retention in the long-term (Fig.\,{\ref{fig:S_Longterm_KCl}}, {\ref{fig:S_Longterm_Compare}}). Note that while the increase in temperature causes a mostly uniform shift in the hysteresis branches as bistability ceases, the experiment involving altered enthalpy primarily affects the dedoping branch. Accordingly, we conclude that the material modification predominantly alters the energetics involved in dedoping, while temperature exerts a more uniform influence on both doping and dedoping.

Normalizing these data sets with respect to $\psi$ and symmetry around $\mu(\psi)=0\,\si{\electronvolt}$ (Fig.\,{\ref{fig:3}}c), we reconstruct the experimental Gibbs free energy profiles by integration and extract the interaction parameters and doping efficiencies with the methods laid out in Supplementary Note 5 and 6. As Fig.\,\ref{fig:3}d and e show, both experiments can be traced back to double-minima Gibbs free energy profiles, where either increasing entropy or decreasing enthalpy cause the local maxima to decrease as anticipated. Remarkably, other works have interpreted the non-volatility of OECTs in a similar, though only conceptual way, without direct evidence of the underlying potential function\autocite{wang2023organic, van2017non}. Here, we can demonstrate and prove for the first time the bistability of an OECT that gives rise to its long-term stable non-volatile behavior. 

Fig.\,\ref{fig:3}f shows the interaction parameters extracted from the data sets of Fig.\,\ref{fig:3}d and e. We find that in all samples, both intraspecies interaction energies ($h_{dd}$ and $h_{uu}$) are of similar small magnitude and for the samples of high enthalpy (green and blue bars), are strongly outweighted by the interspecies parameter $h_{du}$. For the deliberately modified sample (orange bar), there is a reduction across all parameters, with the most notable decrease in $h_{du}$, underscoring the effective suppression of the interactions between doped and undoped sites. Given these parameters, we can further confirm that both bistable systems fulfill the condition of Eq.\,\ref{eq:Condition_Bistability} with $\lambda>1$, which decreases with rising temperature and does not apply to the sample of lowered enthalpy, where we find $\lambda<1$ (Table\,\ref{tab:S_Fitting_alpha}). These findings reinforce the understanding that the bistability arises from the balance between entropic and enthalpic interactions involved in the doping cycle. We want to stress though, that the interaction parameters as shown here yet contain another quantity $Z$, which is the coordination number originating from the derivation of the model and which relates to the microstructure of the assumed doping units (\hyperref[Note_S:Theory]{Supplementary Note 1}). While $Z$ itself is unknown, it is expected a constant in the range typical for three-dimensional structures (i.e., 2 to $\sim8$) and therefore will not change the relative weight of the parameters. 

Regarding their physical significance, the $h$-parameters measure the interaction strength of the doping units among and with each other during the doping cycle. In that context, we posit $h_{du}$ to be coupled to the macroscopic transport properties of the semiconductor film, in line with previous reports highlighting the critical role of domain blending and interaction\autocite{rivnay2016structural,keene2023hole}. It is reasonable to assume that this blending improves with rising temperature, from which a decreased impact of $h_{du}$ would follow, consistent with the framework presented here (Fig.\,\ref{fig:3}d). Similarly, we observe this parameter to decrease with electrostatic screening (Fig.\,\ref{fig:3}e). We interpret this finding to reflect the situation of OECTs measured with aqueous electrolytes, where typically no bistability is observed. Complementing this, Ji et al. have shown the significant enhancement of hysteresis by adding low-polar polytetrahydrofuran (PTHF) to a poly(3,4-ethylenedioxythiophene):tosylate (PEDOT:Tos) channel\autocite{ji2021mimicking}. This finding readily aligns with our model and can be interpreted as follows: Given the aqueous NaCl electrolyte used in their study, a high water content can be expected in the channel, causing little hysteresis in the pristine system due to dielectric screening. That is, enthalpy is suppressed and the system is ruled by entropy. By adding PTHF to the channel, a highly hydrophobic component is introduced, which reduces the exposure to water. It follows a lowered dielectric screening and a rising impact of enthalpy, which gives rise to a distinctive bistability. In this sense, their study appears as a direct counterpart to our experiment of Fig.\,\ref{fig:3}b. Worth highlighting, the authors' microscopic identification of high- and low-resistive regimes involved in charge trapping is a resemblance of the phase transition our model proposes. We expect a similar manifestation for the system studied herein, presumably resulting from the specific interaction between PEDOT:PSS and the ionic liquid{\autocite{kee2016controlling, taussig2024electrostatic, xiong2024counterion}}. In fact, we suspect this phenomenon to extend to a range of organic, non-volatile systems documented in literature{\autocite{ji2021mimicking, wang2023organic, chen2022highly, melianas2020temperature, choi2020vertical}}, under the notion that, despite miscellaneous and system-specific interactions, they share in common that their resulting enthalpic forces exceed entropy.

Our second approach regards the Maxwell construction of Fig.\,\ref{fig:2}b, where the non-monotonic chemical potential is expected to cause a heat exchange. We validate this hypothesis through in-operando thermography studies, which are, to the best of our knowledge, the first of their kind performed at an OECT. As shown in Fig.\,\ref{fig:3}g, we find a heat input equivalent to $1.57\,\si{\kelvin}$ in temperature difference when switching the device on. Switching off again, we note a decreased flux, indicating slower kinetics for dedoping. These findings confirm our hypothesis of an underlying cycle process, where the performance of chemical work is associated with a heat flux changing sign between the two sweeps. At the same time, however, it is obvious that the OECT's transfer curves lack the sharp transition expected from Fig.\,\ref{fig:2}b. This divergence can be attributed to two factors: First, unlike the idealized two-reservoir model, the actual system is not isolated but in thermal contact with its environment. Second, the voltage across the channel is not constant and contributes to the slope of the transition, as shown in \hyperref[Note_S:DrainVoltage]{Supplementary Note 6}. Lastly, we want to stress that for the time being, it is challenging to derive further quantities from this experiment, as the precise energetics of the doping reaction underlying this particular OECT system are unknown. While Rebetez et al. have examined the thermodynamics of the doping process of PEDOT:PSS, their study refers to an aqueous electrolyte and should thus not be transferred to the present system without further ado\autocite{rebetez2022drives}. Nonetheless, our experiment clearly confirms the anticipated heat flux, attributable to the non-monotonic chemical potential.

The third hypothesis finally addresses the subthreshold behavior. We propose that a bistable Gibbs free energy function would manifest itself in a non-monotonic progression of the subthreshold swing with temperature, driven by the balance between enthalpy and entropy (Fig.\,\ref{fig:2}c). In fact, we can prove this exceptional phenomenon experimentally as shown in Fig.\,\ref{fig:3}h (left panel), where the tipping point can be identified at around $30\,\si{\celsius}$. Expectedly, this trend is accompanied by a continuous decrease in hysteresis, while the transconductance shows a similar non-monotonic progression as the subthreshold swing (Fig.\,\ref{fig:NDC_old}). We can further confirm the conjecture of a decrease under reduced enthalpy (Fig.\,\ref{fig:3}h, right panel). As in the former case, this change is driven by a reduced bistability, from which a less pronounced counterbalance from the non-monotonic chemical potential towards the electrostatic potential follows.

\subsection*{Single-OECT Schmitt Trigger}
In view of these results, we consider the bistability for the given OECT system as sufficiently substantiated. This finding unlocks a range of promising applications, in particular for the purpose of neuromorphic computing, where the exploitation of hysteresis as a memory function has already been demonstrated thoroughly\autocite{ji2021mimicking, van2017non, wang2023organic}. Considering its extent (Fig.\,\ref{fig:1}f), the bistability also appears attractive for another application fundamental in digital and auspicious to neuromorphic electronics: the Schmitt trigger\autocite{hopfield1982neural, bharitkar2000hysteretic}. Schmitt triggers are hysteretic devices with separate threshold voltages for rising and falling input. Since the upper and lower output states are only obtained by exceeding these thresholds, suppressive zones are created, making such devices effective noise filters and one-bit analog-to-digital converters (ADCs)\autocite{hwang2020all, bubel2015schmitt}. In neuromorphic computing, such systems can similarly be leveraged to mimic the behavior of biological neurons, which fire only upon exceeding a certain input threshold\autocite{sangwan2020neuromorphic, friedman2016bayesian, zhang2020schmitt, zhang2018coupled}. Conventionally, however, Schmitt triggers are implemented via comparator circuits, often including at least two transistors and six resistors. 

\begin{figure}[t!]
    \centering
    \includegraphics[width=\linewidth]{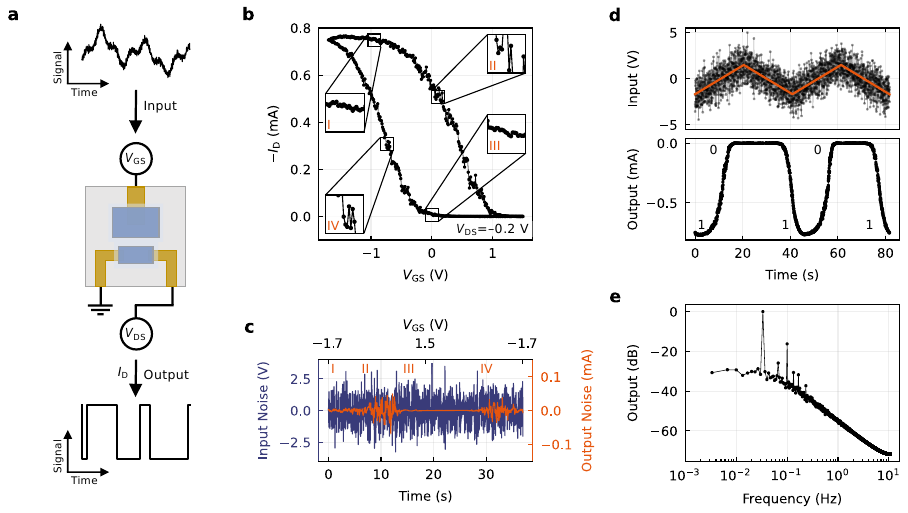} 
    \caption{\textbf{Single-OECT Schmitt trigger.} (\textbf{a}) The bistable OECT functions as a Schmitt trigger, allowing for noise suppression and analog-to-digital conversion. (\textbf{b}) Transfer curve of an OECT ($V_\mathrm{DS}=-0.2\,\si{\volt}$) with offset-shifted white noise input signal (Gaussian noise with $f=400\,\si{\hertz}, V_\mathrm{pp}=10\,\si{\volt}$). Insets: Four distinct regimes of output noise. (\textbf{c}) Input vs. output noise as extracted from (\textbf{b}) against an analytical reference (Fig.\,\ref{fig:S_Schmitt Trigger}). (\textbf{d}) The device functions as an ADC despite severe input noise. (\textbf{e}) Fourier transformation of a first-order Butterworth filter revealing higher harmonics in the output.\label{fig:4}}
     \vspace*{-\baselineskip}
\end{figure}

We explore the applicability of the bistable OECT as a single-device Schmitt trigger, using gate and drain as in- and output (Fig.\,\ref{fig:4}a). Shifting a white noise signal in offset for input, the device shows the expected transfer response, while also clearly reflecting the input fluctuations by corresponding output swings (Fig.\,\ref{fig:4}b). Crucially, however, we can identify four distinct regimes of how the noise translates to the output. During the transitions between the on- and off-state, the noise merges noticeably into the output signal (II and IV), whereas in the sections before transition, there is significant suppression of the same (I and III). In the context of our framework, these transition points appear as being defined by the inflection points $\psi_i$ and $1-\psi_i$, where $\mu(\psi)$ tips into its inverse slope. There, the physical system enters its unstable regime, causing the transfer curve to no longer be determined by the chemical potential alone, and consequently making the output signal more susceptible to a fluctuating input. These regions can be seen even more clearly in Fig.\,\ref{fig:4}c, where the noise has been extracted against an analytical reference (Fig.\,\ref{fig:S_Schmitt Trigger}). The even stronger suppression in III compared to I can thereby be attributed to the off-state being susceptible to only one direction of input deflection. 

With these results already demonstrating the applicability as a noise filter, the time-resolution of the output further demonstrates the use case as an ADC. Taking a noisy triangular function as the input (Fig.\,\ref{fig:4}d, upper panel), the non-steady response allows conversion to bits within a short input window upon exceeding the threshold (Fig.\,\ref{fig:4}d, lower panel). The output states 0 and 1 are each retained over an extended input window despite severe noise. The temporal asymmetry of these states is thereby due to the asymmetric input, to which we bound ourselves for not exhausting the electrochemical window given. 

Finally, we analyze the output in the frequency domain by performing the Fourier transformation of an extended switching cycle. Given an underlying bistable potential function, a periodic perturbation is expected to induce nonlinear oscillations that manifest as higher harmonics (\hyperref[Note_S:Schmitt_Trigger]{Supplementary Note 7}). We can confirm that this is the case for the OECT system as well. The fundamental and odd harmonics are clearly visible as well as several even harmonics with lower amplitude, caused by the non-ideal rectangular shape of the lower half-wave of the output signal. Hence, the signal reveals multiple underlying frequencies, indicating oscillations as expected for a bistable system. To single out one distinct oscillation frequency, an analogue or digital filter can be utilized, e.g., a first-order Butterworth filter, of which the spectrum is shown in Fig.\,\ref{fig:4}e. 

Taking these results together, it is evident that the OECT's intrinsic bistability allows it to operate as a Schmitt trigger, merging the functionality of a multi-component circuit into a single device. As in other cases\autocite{sarkar2022organic, Harikesh2023NatMat}, this is permitted by the key difference between organic electrochemical and conventional devices, namely the use of different charge carrier types and phase systems, inherently allowing for complex operation mechanisms on multiple time scales. 

\section*{Discussion}
In conclusion, we demonstrate, model, and harness bistable OECTs. We set out with a thermodynamic framework, from which we derive the occurrence of bistability as the consequence of enthalpic effects dominating over entropy. We deduce the consequences this situation would entail and prove these experimentally at an appropriately chosen material system. Among others, we show the exceptional scenario in which the subthreshold swing deviates from Boltzmann statistics. Building on these findings, we demonstrate the functionality of a multi-component circuit with a single OECT device, providing a vital building block for advanced organic circuitry. These insights significantly enhance the understanding of OECT physics and set the stage for their application in non-conventional, complex computing, where bistable systems bear enormous upside\autocite{yang2013memristive, boahen2022dendrocentric, talin2022ecram}.

\section*{Methods}

\noindent \textbf{Device fabrication.} Microfabrication of the OECTs was performed on $1\,\si{inch}\times1\,\si{inch}$ glass substrates, onto which Cr ($3\,\si{\nano\metre}$) and Au ($50\,\si{\nano\metre}$) were evaporated for metal traces. AZ 1518 photoresist (MicroChemicals GmbH) was spincoated at 3000\,rpm for $60\,\si{\second}$ (SAWATEC AG spincoating system), followed by baking the substrate at $110\,\si{\celsius}$ for $60\,\si{\second}$. Metal traces were shaped by illuminating in a maskaligner photolithography system (I-line $365\,\si{\nano\metre}$, lamp power $167\,\si{\watt}$, SÜSS Microtec AG) for $10\,\si{\second}$, followed by developing the photoresist in AZ 726 MIF developer (MicroChemicals GmbH), and etching Au and Cr for $60\,\si{\second}$ and $20\,\si{\second}$ (10\% diluted aqueous solutions of Standard Gold/Chromium etchants, Merck KGaA). After $\ce{O2}$-plasma cleaning, PEDOT:PSS (Clevios PH1000, Heraeus Deutschland GmbH \& Co. KG) with 5\%\,v/v ethylene glycol (Merck KGaA) was spincoated (3000\,rpm for $60\,\si{\second}$) to yield a PEDOT:PSS layer of $\sim100\,\si{\nano\metre}$. This step was followed by drying at $120\,\si{\celsius}$ for $20\,\si{\minute}$. Orthogonal photoresist OSCoR 5001 (Orthogonal Inc.) was spincoated (3000\,rpm for $60\,\si{\second}$), baked for $60\,\si{\second}$ at $100\,\si{\celsius}$, and exposed for $12\,\si{\second}$ to structure channel and gate. After post-baking ($60\,\si{\second}$ at $100\,\si{\celsius}$), development followed by covering the sample with Orthogonal Developer 103a (Orthogonal Inc.) and removing it by spinning after $60\,\si{\second}$, which was carried out twice. Excess PEDOT:PSS was removed by $\ce{O2}$-plasma etching for $5\,\si{\minute}$ with $0.3\,\si{\milli\bar}$ (Diener electronic GmbH \& Co. KG). Afterwards, the sample was placed in Orthogonal Stripper 900 (Orthogonal Inc.) over night at room temperature, which was followed by ultrasonic cleaning in ethanol for $15\,\si{\minute}$. Devices in this work had channel dimensions of $L=30\,\si{\micro\metre}$, $W=150\,\si{\micro\metre}$ and featured a PEDOT:PSS side-gate at a distance of $60\,\si{\micro\metre}$. For practical reasons, the thermography and Ag/AgCl-gating experiments were carried out with devices of $L=300\,\si{\micro\metre}$, and $W=150\,\si{\micro\metre}$. Micrographs are provided in Fig.\,{\ref{fig:S_OECT_Figures}}.
\newline \noindent \textbf{Electrolyte preparation and deposition.} Preparation and deposition of the solid-state electrolyte was carried out as put forth in Ref.\,{\citen{weissbach2022photopatternable}}. The precursor solution was prepared by mixing deionized water ($1.0\,\si{\milli\liter}$), N-isopropylacrylamide ($750.0\,\si{\milli\gram}$, Alfa Aesar), N,N'-methylenebisacrylamide ($20.0\,\si{\milli\gram}$, Merck KGaA), 2-hydroxy-4'-(2-hydroxyethoxy)-2-methylpropiophenone ($200.0\,\si{\milli\gram}$, Merck KGaA), and the ionic liquid 1-ethyl-3-methylimidazolium ethyl sulfate ($1.5\,\si{\milli\liter}$, Merck KGaA), followed by stirring over night at room temperature. Before applying the electrolyte, samples were immersed in a 5\%\,v/v solution of 3-(trimethoxysilyl)propyl methacrylate in buffered ethanol (10\%\,v/v acetic acid/acetate) at $50\,\si{\celsius}$ for $10\,\si{\minute}$, in order to deposit an adhesion promoting layer. The sample was thoroughly cleaned with ethanol afterwards and dried at $100\,\si{\celsius}$ for $15\,\si{\minute}$. Deposition of the electrolyte then took place in the maskaligner photolithography system. A drop of precursor solution was put on the devices and carefully covered with a Teflon\textsuperscript{\texttrademark} foil, upon which solidification followed by exposing the areas of interest for $20\,\si{\second}$ through a photomask. Non-crosslinked material was subsequently removed by blowing with a $\ce{N2}$-gun. The electrolyte of the enthalpy-modified systems was deposited by inkjet printing, where the precursor solution was supplemented with $1\,\si{\milli\litre}$ of ethylene glycol and 2 drops of Triton X-100 to set the viscosity. For modifying the enthalpic contribution, $1\,\si{\milli\litre}$ of aqueous KCl solution was added. Before application, filtration was carried out through a $0.45\,\si{\micro\metre}$ PVDF-filter. Placement on the substrate was carried out with a Dimatix Materials DMP-2800 Inkjet Printer, each droplet having a volume of $2.4\,\si{\pico\litre}$ and a spacing of $15\,\si{\micro\metre}$. Solidification took place under UV light for $120\,\si{\second}$.
\newline \noindent \textbf{Electrical characterization.} 
Electrical characterizations were performed in a N$_2$-filled glovebox, with devices sufficiently purged to remove traces of water. Data was acquired using two Keithley 236 SMUs. Temperature-dependent measurements were performed in a N$_2$-filled Janis ST-500 Probe Station with continuous flow cryostat (LN$_2$ cooling), connected to a Scientific Instruments Model 9700 temperature controller and an Agilent HP 4145B. For Schmitt-trigger experiments, a white noise signal was applied through an Agilent 33600A waveform generator. Any setup was controlled by the software SweepMe! (sweep-me.net). All transfer curves were recorded at sufficiently low scan rates ($\sim10^{-2}\,\si{\volt\per\second}$) to minimize kinetic effects.
\newline \noindent \textbf{Thermography measurements.} Thermography measurements were performed using an InfraTec ImageIR 9420 thermal camera, installed on a semiautomatic wafer probe station Formfactor CM300. Thus it was possible to align, contact, and measure the samples electrically using single probes while recording the infrared image simultaneously. Electrical biasing and measurements were done via a Keysight B1500 semiconductor parameter analyzer.
\newline \noindent \textbf{Fits and simulations.} Fits and simulations were carried out with customised Python scripts. Gibbs free energy functions were reconstructed by the methods laid out in \hyperref[Note_S:G_Fitting]{Supplementary Note 5}.

\section*{Data Availability}

The data underlying the figures in the main text are publicly available from the OPARA repository at \url{https://doi.org/10.25532/OPARA-552}. The datasets generated and/or analyzed during the study are available from the corresponding author upon request.

\section*{Code Availability}

The code underlying the free energy fits is publicly available at \url{https://github.com/lukasbongartz/thermodynamics-fitting.git}.

\section*{References}
\printbibliography[heading = none]

@article{rivnay2018organic,
  title={{Organic electrochemical transistors}},
  author={Rivnay, Jonathan and Inal, Sahika and Salleo, Alberto and Owens, R{\'o}is{\'\i}n M and Berggren, Magnus and Malliaras, George G},
  journal={Nature Reviews Materials},
  volume={3},
  number={2},
  pages={1--14},
  year={2018},
  publisher={Nature Publishing Group}
}

@article{gkoupidenis2015neuromorphic,
  title={{Neuromorphic Functions in {PEDOT:PSS} Organic Electrochemical Transistors}},
  author={Gkoupidenis, Paschalis and Schaefer, Nathan and Garlan, Benjamin and Malliaras, George G},
  journal={Advanced Materials},
  volume={27},
  number={44},
  pages={7176--7180},
  year={2015},
  publisher={Wiley Online Library}
}

@article{friedlein2018device,
  title={Device physics of organic electrochemical transistors},
  author={Friedlein, Jacob T and McLeod, Robert R and Rivnay, Jonathan},
  journal={Organic Electronics},
  volume={63},
  pages={398--414},
  year={2018},
  publisher={Elsevier}
}

@article{van2018organic,
  title={Organic electronics for neuromorphic computing},
  author={van De Burgt, Yoeri and Melianas, Armantas and Keene, Scott Tom and Malliaras, George and Salleo, Alberto},
  journal={Nature Electronics},
  volume={1},
  number={7},
  pages={386--397},
  year={2018},
  publisher={Nature Publishing Group}
}

@article{rivnay2016structural,
  title={Structural control of mixed ionic and electronic transport in conducting polymers},
  author={Rivnay, Jonathan and Inal, Sahika and Collins, Brian A and Sessolo, Michele and Stavrinidou, Eleni and Strakosas, Xenofon and Tassone, Christopher and Delongchamp, Dean M and Malliaras, George G},
  journal={Nature Communications},
  volume={7},
  number={1},
  pages={11287},
  year={2016},
  publisher={Nature Publishing Group UK London}
}

@article{weissbach2022photopatternable,
  title={Photopatternable solid electrolyte for integrable organic electrochemical transistors: operation and hysteresis},
  author={Weissbach, Anton and Bongartz, Lukas M and Cucchi, Matteo and Tseng, Hsin and Leo, Karl and Kleemann, Hans},
  journal={Journal of Materials Chemistry C},
  volume={10},
  number={7},
  pages={2656--2662},
  year={2022},
  publisher={Royal Society of Chemistry}
}

@article{ji2021mimicking,
  title={Mimicking associative learning using an ion-trapping non-volatile synaptic organic electrochemical transistor},
  author={Ji, Xudong and Paulsen, Bryan D and Chik, Gary KK and Wu, Ruiheng and Yin, Yuyang and Chan, Paddy KL and Rivnay, Jonathan},
  journal={Nature Communications},
  volume={12},
  number={1},
  pages={1--12},
  year={2021},
  publisher={Nature Publishing Group}
}

@article{rebetez2022drives,
  title={What drives the kinetics and doping level in the electrochemical reactions of {PEDOT: PSS}?},
  author={Rebetez, Gonzague and Bardagot, Olivier and Affolter, Jo{\"e}l and R{\'e}hault, Julien and Banerji, Natalie},
  journal={Advanced Functional Materials},
  volume={32},
  number={5},
  pages={2105821},
  year={2022},
  publisher={Wiley Online Library}
}

@article{paulsen2020organic,
  title={Organic mixed ionic--electronic conductors},
  author={Paulsen, Bryan D and Tybrandt, Klas and Stavrinidou, Eleni and Rivnay, Jonathan},
  journal={Nature Materials},
  volume={19},
  number={1},
  pages={13--26},
  year={2020},
  publisher={Nature Publishing Group}
}

@article{van2017non,
  title={A non-volatile organic electrochemical device as a low-voltage artificial synapse for neuromorphic computing},
  author={van De Burgt, Yoeri and Lubberman, Ewout and Fuller, Elliot J and Keene, Scott T and Faria, Gr{\'e}gorio C and Agarwal, Sapan and Marinella, Matthew J and Alec Talin, A and Salleo, Alberto},
  journal={Nature Materials},
  volume={16},
  number={4},
  pages={414--418},
  year={2017},
  publisher={Nature Publishing Group}
}

@article{bernards2007steady,
  title={Steady-state and transient behavior of organic electrochemical transistors},
  author={Bernards, Daniel A and Malliaras, George G},
  journal={Advanced Functional Materials},
  volume={17},
  number={17},
  pages={3538--3544},
  year={2007},
  publisher={Wiley Online Library}
}

@inproceedings{bongartz2022temperature,
  title={{Temperature-Dependence of All-Solid-State Organic Electrochemical Transistors}},
  author={Bongartz, Lukas M and Weissbach, Anton and Cucchi, Matteo and Leo, Karl and Kleemann, Hans},
  booktitle={2022 IEEE International Conference on Flexible and Printable Sensors and Systems (FLEPS)},
  pages={1--4},
  year={2022},
  organization={IEEE}
}

@article{cucchi2022thermodynamics,
  title={Thermodynamics of organic electrochemical transistors},
  author={Cucchi, Matteo and Weissbach, Anton and Bongartz, Lukas M and Kantelberg, Richard and Tseng, Hsin and Kleemann, Hans and Leo, Karl},
  journal={Nature Communications},
  volume={13},
  number={1},
  pages={4514},
  year={2022},
  publisher={Nature Publishing Group UK London}
}

@article{dreyer2010thermodynamic,
  title={The thermodynamic origin of hysteresis in insertion batteries},
  author={Dreyer, Wolfgang and Jamnik, Janko and Guhlke, Clemens and Huth, Robert and Mo{\v{s}}kon, Jo{\v{z}}e and Gaber{\v{s}}{\v{c}}ek, Miran},
  journal={Nature Materials},
  volume={9},
  number={5},
  pages={448--453},
  year={2010},
  publisher={Nature Publishing Group}
}

@article{kaphle2018organic,
  title={Organic electrochemical transistors based on room temperature ionic liquids: performance and stability},
  author={Kaphle, Vikash and Liu, Shiyi and Keum, Chang-Min and L{\"u}ssem, Bj{\"o}rn},
  journal={Pysica Status Solidi A},
  volume={215},
  number={24},
  pages={1800631},
  year={2018},
  publisher={Wiley Online Library}
}

@article{sarkar2022organic,
  title={An organic artificial spiking neuron for in situ neuromorphic sensing and biointerfacing},
  author={Sarkar, Tanmoy and Lieberth, Katharina and Pavlou, Aristea and Frank, Thomas and Mailaender, Volker and McCulloch, Iain and Blom, Paul WM and Torricelli, Fabrizio and Gkoupidenis, Paschalis},
  journal={Nature Electronics},
  volume={5},
  number={11},
  pages={774--783},
  year={2022},
  publisher={Nature Publishing Group}
}

@article{bubel2015schmitt,
  title={Schmitt trigger using a self-healing ionic liquid gated transistor},
  author={Bubel, Simon and Menyo, Matthew S and Mates, Thomas E and Waite, J Herbert and Chabinyc, Michael L},
  journal={Advanced Materials},
  volume={27},
  number={21},
  pages={3331--3335},
  year={2015},
  publisher={Wiley Online Library}
}

@article{zhang2020schmitt,
  title={{A Schmitt Trigger Based Oscillatory Neural Network for Reservoir Computing}},
  author={Zhang, Ting and Haider, Mohammad R},
  journal={Journal of Electrical and Electronic Engineering},
  volume={8},
  pages={1--9},
  year={2020}
}

@inproceedings{zhang2018coupled,
  author={Zhang, Ting and Haider, Mohammad R. and Alexander, Iwan D. and Massoud, Yehia},
  booktitle={2018 IEEE 61st International Midwest Symposium on Circuits and Systems (MWSCAS)}, 
  title={A Coupled Schmitt Trigger Oscillator Neural Network for Pattern Recognition Applications}, 
  year={2018},
  volume={},
  number={},
  pages={238-241},
  doi={10.1109/MWSCAS.2018.8624010}
}

@article{hwang2020all,
  title={{All-Solid-State Organic Schmitt Trigger Implemented by Twin Two-in-One Ferroelectric Memory Transistors}},
  author={Hwang, Sunbin and Jang, Sukjae and Bae, Sukang and Lee, Seoung-Ki and Lee, Sang Hyun and Fabiano, Simone and Berggren, Magnus and Lee, Takhee and Kim, Tae-Wook},
  journal={Advanced Electronic Materials},
  volume={6},
  number={5},
  pages={1901263},
  year={2020},
  publisher={Wiley Online Library}
}

@article{hopfield1982neural,
  title={Neural networks and physical systems with emergent collective computational abilities.},
  author={Hopfield, John J},
  journal={Proceedings of the National Academy of Sciences},
  volume={79},
  number={8},
  pages={2554--2558},
  year={1982},
  publisher={National Acad Sciences}
}

@article{bharitkar2000hysteretic,
  title={The hysteretic Hopfield neural network},
  author={Bharitkar, Sunil and Mendel, Jerry M},
  journal={IEEE Transactions on neural networks},
  volume={11},
  number={4},
  pages={879--888},
  year={2000},
  publisher={IEEE}
}

@article{friedman2016bayesian,
  title={Bayesian inference with {M}uller {C}-elements},
  author={Friedman, Joseph S and Calvet, Laurie E and Bessi{\`e}re, Pierre and Droulez, Jacques and Querlioz, Damien},
  journal={IEEE Transactions on Circuits and Systems I: Regular Papers},
  volume={63},
  number={6},
  pages={895--904},
  year={2016},
  publisher={IEEE}
}

@article{Harikesh2023NatMat,
  title={Ion-tunable antiambipolarity in mixed ion--electron conducting polymers enables biorealistic organic electrochemical neurons},
  author={Harikesh, Padinhare Cholakkal and Yang, Chi-Yuan and Wu, Han-Yan and Zhang, Silan and Donahue, Mary J and Caravaca, April S and Huang, Jun-Da and Olofsson, Peder S and Berggren, Magnus and Tu, Deyu and others},
  journal={Nature Materials},
  volume={22},
  number={2},
  pages={242--248},
  year={2023},
  publisher={Nature Publishing Group UK London}
}

@article{sangwan2020neuromorphic,
  title={Neuromorphic nanoelectronic materials},
  author={Sangwan, Vinod K and Hersam, Mark C},
  journal={Nature Nanotechnology},
  volume={15},
  number={7},
  pages={517--528},
  year={2020},
  publisher={Nature Publishing Group}
}

@article{ohayon2023guide,
  title={A guide for the characterization of organic electrochemical transistors and channel materials},
  author={Ohayon, David and Druet, Victor and Inal, Sahika},
  journal={Chemical Society Reviews},
  volume={52},
  number={3},
  pages={1001--1023},
  year={2023},
  publisher={Royal Society of Chemistry}
}

@article{huang2023vertical,
  title={Vertical organic electrochemical transistors for complementary circuits},
  author={Huang, Wei and Chen, Jianhua and Yao, Yao and Zheng, Ding and Ji, Xudong and Feng, Liang-Wen and Moore, David and Glavin, Nicholas R and Xie, Miao and Chen, Yao and others},
  journal={Nature},
  volume={613},
  number={7944},
  pages={496--502},
  year={2023},
  publisher={Nature Publishing Group UK London}
}

@article{yang2013memristive,
  title={Memristive devices for computing},
  author={Yang, J Joshua and Strukov, Dmitri B and Stewart, Duncan R},
  journal={Nature Nanotechnology},
  volume={8},
  number={1},
  pages={13--24},
  year={2013},
  publisher={Nature Publishing Group UK London}
}

@article{jing2017hygroscopic,
  title={Hygroscopic properties of potassium chloride and its internal mixtures with organic compounds relevant to biomass burning aerosol particles},
  author={Jing, Bo and Peng, Chao and Wang, Yidan and Liu, Qifan and Tong, Shengrui and Zhang, Yunhong and Ge, Maofa},
  journal={Scientific Reports},
  volume={7},
  number={1},
  pages={43572},
  year={2017},
  publisher={Nature Publishing Group UK London}
}

@article{boahen2022dendrocentric,
  title={Dendrocentric learning for synthetic intelligence},
  author={Boahen, Kwabena},
  journal={Nature},
  volume={612},
  number={7938},
  pages={43--50},
  year={2022},
  publisher={Nature Publishing Group UK London}
}

@book{borgnakke2022fundamentals,
  title={Fundamentals of Thermodynamics},
  author={Borgnakke, Claus and Sonntag, Richard Edwin},
  year={2022},
  publisher={John Wiley \& Sons}
}

@article{shameem2022hysteresis,
  title={{Hysteresis in Organic Electrochemical Transistors: Relation to the Electrochemical Properties of the Semiconductor}},
  author={Shameem, Raufar and Bongartz, Lukas M and Weissbach, Anton and Kleemann, Hans and Leo, Karl},
  journal={Applied Sciences},
  volume={13},
  number={9},
  pages={5754},
  year={2023},
  publisher={MDPI}
}

@article{wang2023organic,
  title={An organic electrochemical transistor for multi-modal sensing, memory and processing},
  author={Wang, Shijie and Chen, Xi and Zhao, Chao and Kong, Yuxin and Lin, Baojun and Wu, Yongyi and Bi, Zhaozhao and Xuan, Ziyi and Li, Tao and Li, Yuxiang and others},
  journal={Nature Electronics},
  volume={6},
  number={4},
  pages={281--291},
  year={2023},
  publisher={Nature Publishing Group UK London}
}

@article{cucchi2023liquido,
  title={In liquido computation with electrochemical transistors and mixed conductors for intelligent bioelectronics},
  author={Cucchi, Matteo and Parker, Daniela and Stavrinidou, Eleni and Gkoupidenis, Paschalis and Kleemann, Hans},
  journal={Advanced Materials},
  volume={35},
  number={15},
  pages={2209516},
  year={2023},
  publisher={Wiley Online Library}
}

@article{someya2016rise,
  title={The rise of plastic bioelectronics},
  author={Someya, Takao and Bao, Zhenan and Malliaras, George G},
  journal={Nature},
  volume={540},
  number={7633},
  pages={379--385},
  year={2016},
  publisher={Nature Publishing Group UK London}
}

@article{park2018self,
  title={Self-powered ultra-flexible electronics via nano-grating-patterned organic photovoltaics},
  author={Park, Sungjun and Heo, Soo Won and Lee, Wonryung and Inoue, Daishi and Jiang, Zhi and Yu, Kilho and Jinno, Hiroaki and Hashizume, Daisuke and Sekino, Masaki and Yokota, Tomoyuki and others},
  journal={Nature},
  volume={561},
  number={7724},
  pages={516--521},
  year={2018},
  publisher={Nature Publishing Group UK London}
}

@article{talin2022ecram,
  title={{{ECRAM} materials, devices, circuits and architectures: A perspective}},
  author={Talin, A Alec and Li, Yiyang and Robinson, Donald A and Fuller, Elliot J and Kumar, Suhas},
  journal={Advanced Materials},
  volume={35},
  number={37},
  pages={2204771},
  year={2023},
  publisher={Wiley Online Library}
}

@article{vela2014key,
  title={The key role of vibrational entropy in the phase transitions of dithiazolyl-based bistable magnetic materials},
  author={Vela, Sergi and Mota, Fernando and Deumal, Merc{\`e} and Suizu, Rie and Shuku, Yoshiaki and Mizuno, Asato and Awaga, Kunio and Shiga, Motoyuki and Novoa, Juan J and Ribas-Arino, Jordi},
  journal={Nature Communications},
  volume={5},
  number={1},
  pages={4411},
  year={2014},
  publisher={Nature Publishing Group UK London}
}

@article{koulakov2002model,
  title={Model for a robust neural integrator},
  author={Koulakov, Alexei A and Raghavachari, Sridhar and Kepecs, Adam and Lisman, John E},
  journal={Nature Neuroscience},
  volume={5},
  number={8},
  pages={775--782},
  year={2002},
  publisher={Nature Publishing Group US New York}
}

@article{keene2023hole,
  title={Hole-limited electrochemical doping in conjugated polymers},
  author={Keene, Scott T and Laulainen, Joonatan EM and Pandya, Raj and Moser, Maximilian and Schnedermann, Christoph and Midgley, Paul A and McCulloch, Iain and Rao, Akshay and Malliaras, George G},
  journal={Nature Materials},
  volume={22},
  number={9},
  pages={1121--1127},
  year={2023},
  publisher={Nature Publishing Group UK London}
}

@article{weingaertner2014static,
  title={The static dielectric permittivity of ionic liquids},
  author={Weingaertner, Hermann},
  journal={Journal of Molecular Liquids},
  volume={192},
  pages={185--190},
  year={2014},
  publisher={Elsevier}
}

@book{sze2021physics,
  title={Physics of Semiconductor Devices},
  author={Sze, Simon M and Li, Yiming and Ng, Kwok K},
  year={2021},
  publisher={John Wiley \& Sons}
}

@article{wu2019ionic,
  title={Ionic-liquid doping enables high transconductance, fast response time, and high ion sensitivity in organic electrochemical transistors},
  author={Wu, Xihu and Surendran, Abhijith and Ko, Jieun and Filonik, Oliver and Herzig, Eva M and M{\"u}ller-Buschbaum, Peter and Leong, Wei Lin},
  journal={Advanced Materials},
  volume={31},
  number={2},
  pages={1805544},
  year={2019},
  publisher={Wiley Online Library}
}

@article{gkoupidenis2023organic,
  title={Organic mixed conductors for bioinspired electronics},
  author={Gkoupidenis, Paschalis and Zhang, Y and Kleemann, H and Ling, H and Santoro, F and Fabiano, Simone and Salleo, A and van de Burgt, Y},
  journal={Nature Reviews Materials},
  volume={9},
  number={2},
  pages={134--149},
  year={2024},
  publisher={Nature Publishing Group UK London}
}

@article{bisquert2023hysteresis,
  title={{Hysteresis in Organic Electrochemical Transistors: Distinction of Capacitive and Inductive Effects}},
  author={Bisquert, Juan},
  journal={The Journal of Physical Chemistry Letters},
  volume={14},
  number={49},
  pages={10951--10958},
  year={2023},
  publisher={ACS Publications}
}

@article{2020SciPy-NMeth,
  author  = {Virtanen, Pauli and Gommers, Ralf and Oliphant, Travis E. and
            Haberland, Matt and Reddy, Tyler and Cournapeau, David and
            Burovski, Evgeni and Peterson, Pearu and Weckesser, Warren and
            Bright, Jonathan and {van der Walt}, St{\'e}fan J. and
            Brett, Matthew and Wilson, Joshua and Millman, K. Jarrod and
            Mayorov, Nikolay and Nelson, Andrew R. J. and Jones, Eric and
            Kern, Robert and Larson, Eric and Carey, C J and
            Polat, {\.I}lhan and Feng, Yu and Moore, Eric W. and
            {VanderPlas}, Jake and Laxalde, Denis and Perktold, Josef and
            Cimrman, Robert and Henriksen, Ian and Quintero, E. A. and
            Harris, Charles R. and Archibald, Anne M. and
            Ribeiro, Ant{\^o}nio H. and Pedregosa, Fabian and
            {van Mulbregt}, Paul and {SciPy 1.0 Contributors}},
  title   = {{{SciPy} 1.0: Fundamental Algorithms for Scientific
            Computing in Python}},
  journal = {Nature Methods},
  year    = {2020},
  volume  = {17},
  pages   = {261--272},
  adsurl  = {https://rdcu.be/b08Wh},
  doi     = {10.1038/s41592-019-0686-2},
}

@misc{bistability_simulation,
  title        = {{Simulation Tool: Bistable Organic Electrochemical Transistors}},
  author       = {Lukas M. Bongartz},
  year         = 2024,
  note         = {\url{https://bit.ly/bistability}}
}

@article{kee2016controlling,
  title={Controlling molecular ordering in aqueous conducting polymers using ionic liquids},
  author={Kee, Seyoung and Kim, Nara and Kim, Bong Seong and Park, Seongjin and Jang, Yun Hee and Lee, Seoung Ho and Kim, Jehan and Kim, Junghwan and Kwon, Sooncheol and Lee, Kwanghee},
  journal={Adv. Mater},
  volume={28},
  number={39},
  pages={8625--8631},
  year={2016}
}

@article{chen2022highly,
  title={Highly stretchable organic electrochemical transistors with strain-resistant performance},
  author={Chen, Jianhua and Huang, Wei and Zheng, Ding and Xie, Zhaoqian and Zhuang, Xinming and Zhao, Dan and Chen, Yao and Su, Ning and Chen, Hongming and Pankow, Robert M and others},
  journal={Nature Materials},
  volume={21},
  number={5},
  pages={564--571},
  year={2022},
  publisher={Nature Publishing Group UK London}
}

@article{taussig2024electrostatic,
  title={Electrostatic self-assembly yields a structurally stabilized PEDOT: PSS with efficient mixed transport and high-performance OECTs},
  author={Taussig, Laine and Ghasemi, Masoud and Han, Sanggil and Kwansa, Albert L and Li, Ruipeng and Keene, Scott T and Woodward, Nathan and Yingling, Yaroslava G and Malliaras, George G and Gomez, Enrique D and others},
  journal={Matter},
  volume={7},
  number={3},
  pages={1071--1091},
  year={2024},
  publisher={Elsevier}
}

@article{melianas2020temperature,
  title={Temperature-resilient solid-state organic artificial synapses for neuromorphic computing},
  author={Melianas, Armantas and Quill, TJ and LeCroy, G and Tuchman, Y and Loo, H v and Keene, ST and Giovannitti, Alexander and Lee, HR and Maria, IP and McCulloch, Iain and others},
  journal={Science advances},
  volume={6},
  number={27},
  pages={eabb2958},
  year={2020},
  publisher={American Association for the Advancement of Science}
}

@article{choi2020vertical,
  title={Vertical organic synapse expandable to 3D crossbar array},
  author={Choi, Yongsuk and Oh, Seyong and Qian, Chuan and Park, Jin-Hong and Cho, Jeong Ho},
  journal={Nature Communications},
  volume={11},
  number={1},
  pages={4595},
  year={2020},
  publisher={Nature Publishing Group UK London}
}

@article{xiong2024counterion,
  title={Counterion docking: a general approach to reducing energetic disorder in doped polymeric semiconductors},
  author={Xiong, Miao and Deng, Xin-Yu and Tian, Shuang-Yan and Liu, Kai-Kai and Fang, Yu-Hui and Wang, Juan-Rong and Wang, Yunfei and Liu, Guangchao and Chen, Jupeng and Villalva, Diego Rosas and others},
  journal={Nature Communications},
  volume={15},
  number={1},
  pages={4972},
  year={2024},
  publisher={Nature Publishing Group UK London}
}

\section*{Acknowledgments}
L.M.B. and H.K. are grateful for funding from the German Research Foundation (DFG) under the grant KL 2961/5-1 and the Bundesministerium f{\"u}r Bildung und Forschung (BMBF) for funding from the project BAYOEN (01IS21089). H.K and C.M. acknowledge the project ArNeBOT funded by the Deutsche Forschungsgemeinschaft (DFG, German Research Foundation) – 536022519. R.K. thanks the Hector Fellow Academy for support. The authors acknowledge funding from the ct.qmat Cluster of Excellence.

\section*{Author Contributions}
L.M.B., M.C., and H.K. developed the concept and derived the math. L.M.B. performed simulations. L.M.B., M.C., H.K. and K.L. designed the experiments. A.W. developed the electrolyte. L.M.B. and T.M. carried out device fabrication and transistor measurements. R.H. performed the thermography studies. R.K. and L.M.B. performed the Gibbs free energy fitting and C.M. computed the Butterworth filter. L.M.B. wrote the manuscript. All authors contributed to conceptual discussions and manuscript editing.

\section*{Competing Interests}

The authors declare no competing interests.

\end{document}

\singlespacing
\raggedbottom
\clearpage

\title[]{Bistable Organic Electrochemical Transistors: Enthalpy vs. Entropy}
\small{\centering
\author{Lukas M. Bongartz$^{1*}$, Richard Kantelberg$^{1}$, Tommy Meier$^{1}$, Raik Hoffmann$^{2}$, Christian Matthus$^{3}$, Anton Weissbach$^{1}$, Matteo Cucchi$^{1}$, Hans Kleemann$^{1}$, Karl Leo$^{1}$}

\address{$^{1}$IAPP Dresden, Institute for Applied Physics, Technische Universit\"at Dresden, N\"othnitzer Str. 61, 01187 Dresden, Germany}
\address{$^{2}$Fraunhofer Institute for Photonic Microsystems IPMS, Center Nanoelectronic Technologies, An der Bartlake 5, 01099 Dresden, Germany}
\address{$^{3}$Chair of Circuit Design and Network Theory (CCN), Faculty of Electrical and Computer Engineering, Technische Universit\"at Dresden, Helmholtzstr. 18, 01069 Dresden, Germany}}

\section*{\fontsize{14pt}{16pt}\selectfont \centering Supplementary Information}

\startcontents[section]
\printcontents[section]{}{1}{\section*{Contents}}

\setcounter{equation}{0}
\renewcommand{\theequation}{S\arabic{equation}}
\setcounter{figure}{0}
\renewcommand{\thefigure}{S\arabic{figure}}
\setcounter{table}{0}
\renewcommand{\thetable}{S\arabic{table}}

\clearpage

\section*{Supplementary Figures}
\addcontentsline{toc}{section}{Supplementary Figures}

\begin{figure}[H]
    \centering
    \includegraphics[width=0.5\linewidth]{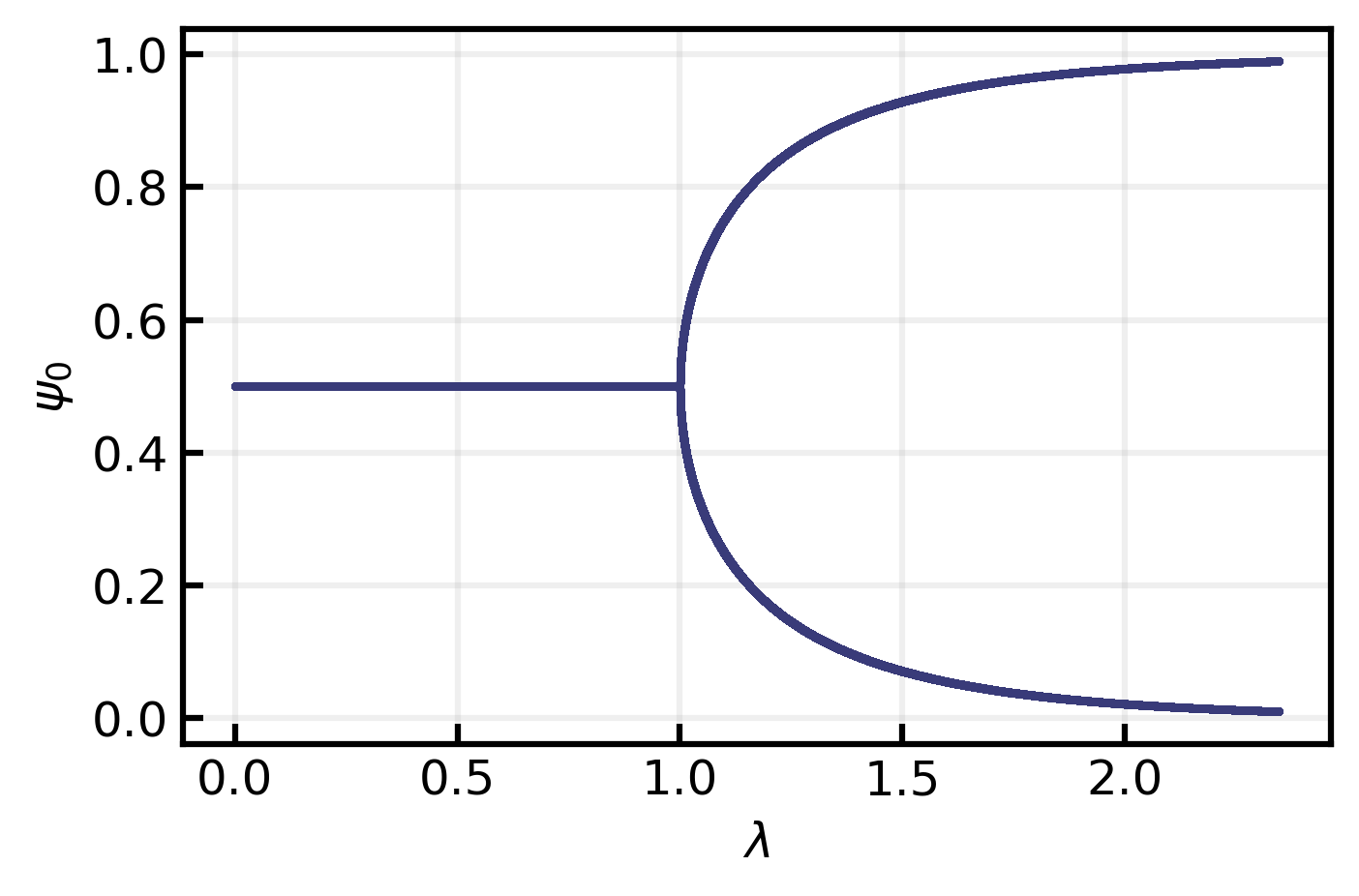} 
    \caption{\textbf{Bifurcation diagram.} For $\lambda>1$ (Eq.\,\ref{eq:Condition_Bistability}), enthalpic contributions to the Gibbs free energy $G(\psi)$ dominate over entropic contributions, which leads to a bifurcation of the equilibrium state at $\psi_0$.  Shown here is the case of $h_{uu}=h_{dd}=0$ for $\psi=0.5$. 
  \label{fig:S_Bifurcation}}
\end{figure}
\addcontentsline{toc}{subsection}{Fig.~\protect\numberline{\thefigure:}Bifurcation diagram}

\begin{figure}[H]
    \centering
    \includegraphics[width=\linewidth]{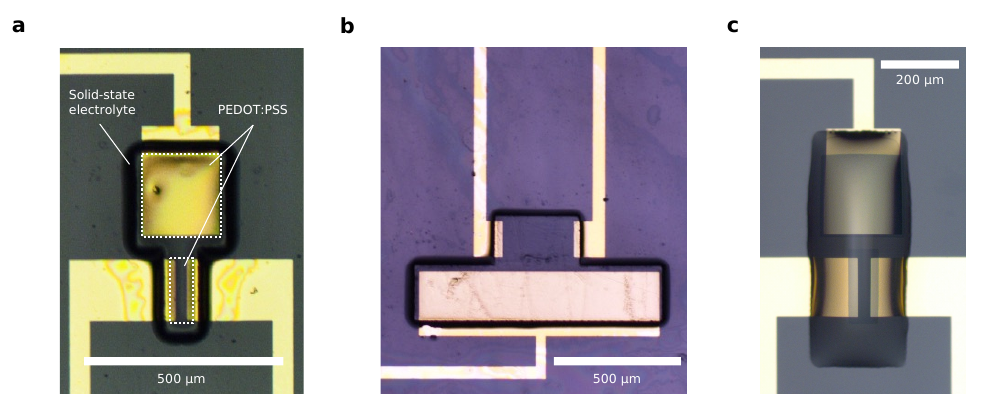} 
    \caption{\textbf{Micrographs of solid-state OECTs.} Lithography-fabricated devices with channel dimensions of (\textbf{a}) $L=30\,\si{\micro\metre}$, $W=150\,\si{\micro\metre}$ and (\textbf{b}) $L=300\,\si{\micro\metre}$, $W=150\,\si{\micro\metre}$. (\textbf{c}) Devices fabricated in a hybrid process (lithography and inkjet-printing) with $L=30\,\si{\micro\metre}$, $W=150\,\si{\micro\metre}$ and modified electrolyte system (hygroscopic KCl additive).\label{fig:S_OECT_Figures}}
\end{figure}
\addcontentsline{toc}{subsection}{Fig.~\protect\numberline{\thefigure:}Micrographs of solid-state OECTs}

\begin{figure}[H]
    \centering
    \includegraphics[width=\linewidth]{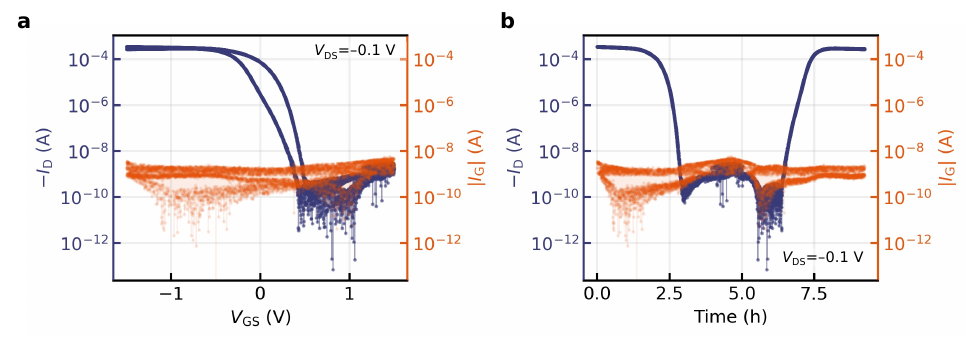} 
    \caption{\textbf{Transfer curve with extremely reduced scan rate.} (a) Transfer curve of the solid-state OECT with a scan rate of $180\,\si{\micro\volt\per\second}$ with (b) the transient response.\label{fig:S_Transfer_Slow}}
\end{figure}
\addcontentsline{toc}{subsection}{Fig.~\protect\numberline{\thefigure:}Transfer curve with extremely reduced scan rate}

\begin{figure}[H]
    \centering
    \includegraphics[width=\linewidth]{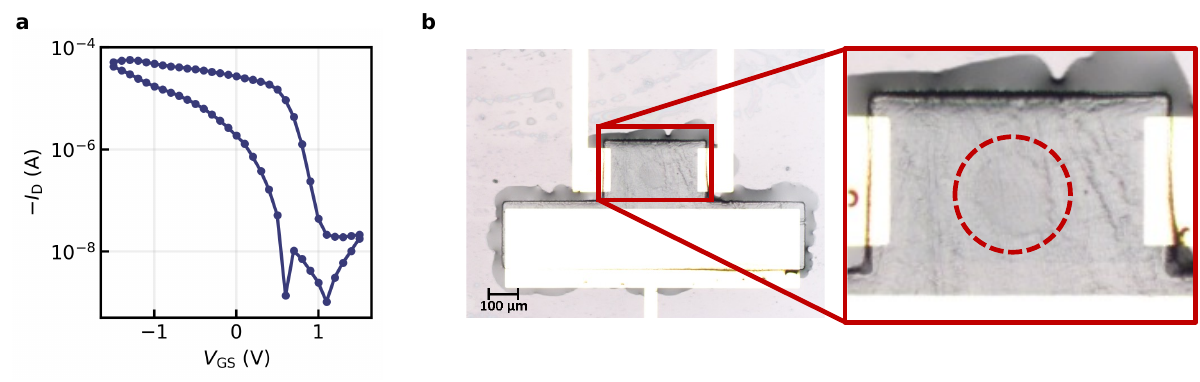} 
    \caption{\textbf{Solid-state OECT with Ag/AgCl gate.} (\textbf{a}) Verification of hysteresis when operating with an Ag/AgCl gate. (\textbf{b}) Imprint of $80\,\si{\micro\metre}$ diameter gate verifies that the Au side-gate was not in contact. \label{fig:S_AgAgClGate}}
\end{figure}
\addcontentsline{toc}{subsection}{Fig.~\protect\numberline{\thefigure:}Solid-state OECT with Ag/AgCl gate}

\begin{figure}[H]
    \centering
    \includegraphics[width=0.7\linewidth]{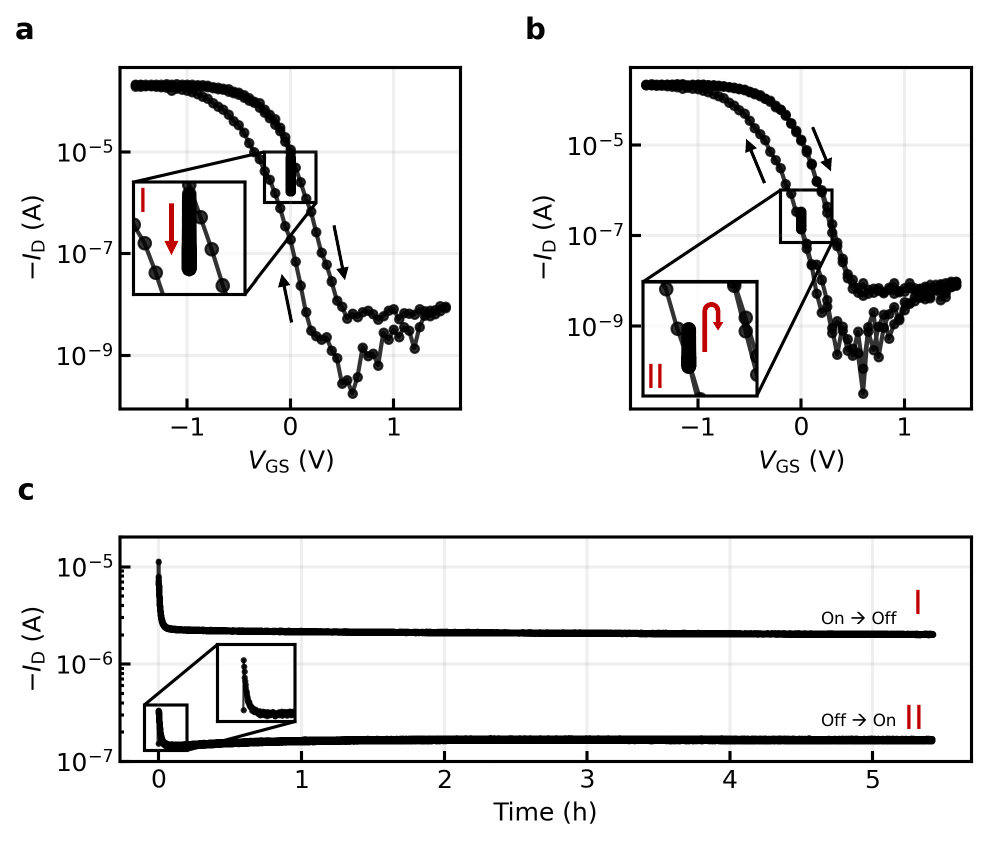} 
    \caption{\textbf{Bistability in time domain.} Operating an OECT and holding a $0\,\si{\volt}$ gate bias approaching from the (\textbf{a}) on- or (\textbf{b}) off-state reveals (\textbf{c}) two coexisting equilibrium states.
    \label{fig:S_Normal_longterm}}
\end{figure}
\addcontentsline{toc}{subsection}{Fig.~\protect\numberline{\thefigure:}Bistability in time domain}

\begin{figure}[H]
    \centering
    \includegraphics[width=0.9\linewidth]{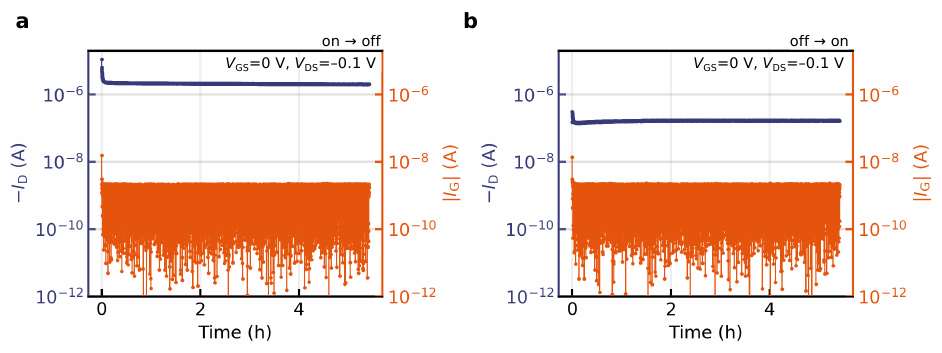} 
    \caption{\textbf{Transient state retention with gate current.} State retention in (a) low-resistance state and (b) high-resistance state including the gate current.\label{fig:S_Normal_longterm_GateCurrent}}
\end{figure}
\addcontentsline{toc}{subsection}{Fig.~\protect\numberline{\thefigure:}Transient state retention with gate current}

\begin{figure}[H]
    \centering
    \includegraphics[width=0.8\textwidth]{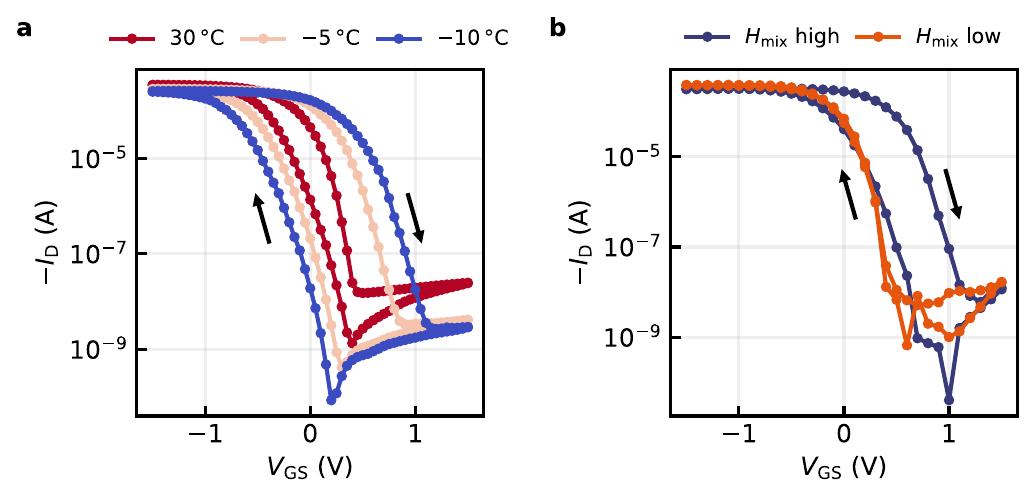}
  \caption{\textbf{Suppressing the bistability via entropy and enthalpy (log scale).} Data of (\textbf{a}) Fig.\,\ref{fig:3}a and (\textbf{b}) Fig.\,\ref{fig:3}b in logarithmic scale. Note also the change in subthreshold swing for both cases. \label{fig:S_KCl_log}}
\end{figure}
\addcontentsline{toc}{subsection}{Fig.~\protect\numberline{\thefigure:}Suppressing the bistability via entropy and enthalpy (log scale)}

\begin{figure}[H]
    \centering
    \includegraphics[width=0.7\linewidth]{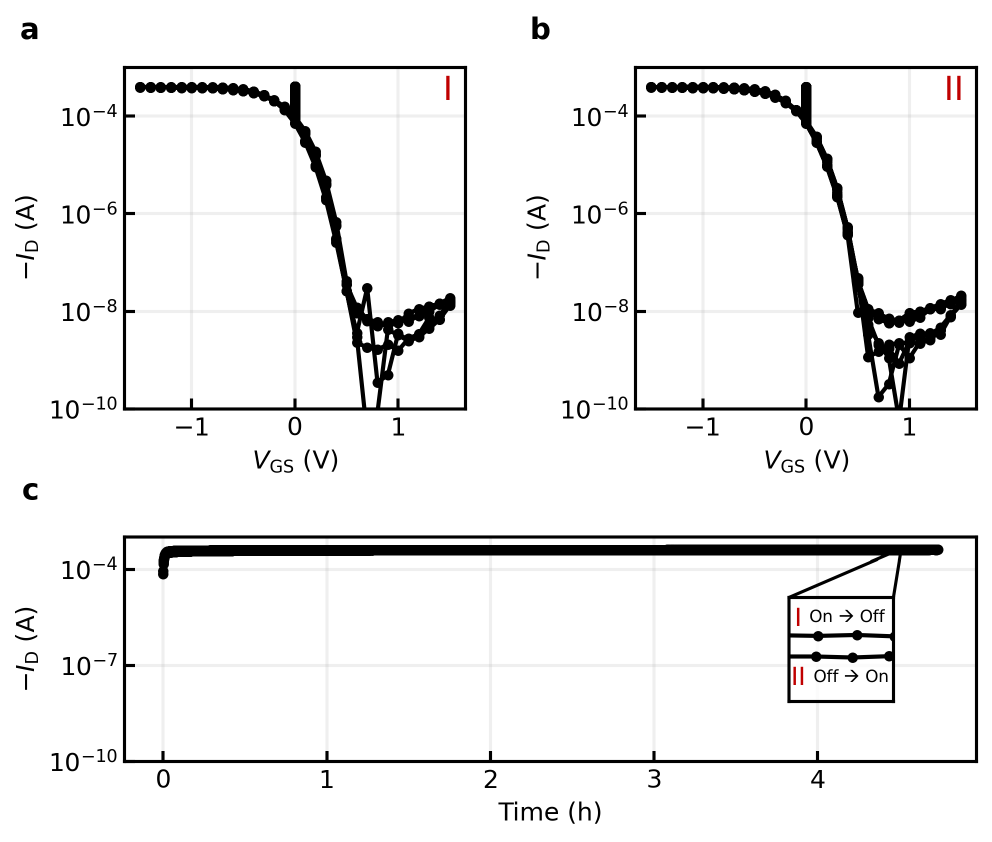} 
    \caption{\textbf{Long-term validation of suppressed bistability.} The experiment of Fig.\,\ref{fig:1}d and \ref{fig:S_Normal_longterm} was carried out with the system of lowered enthalpy (hygroscopic KCl additive). (\textbf{a}) Approaching from the on-state. (\textbf{b}) Approaching from the off-state. (\textbf{c}) Both tracks approach the same current level, confirming the transition from bi- to monostability.\label{fig:S_Longterm_KCl}}
\end{figure}
\addcontentsline{toc}{subsection}{Fig.~\protect\numberline{\thefigure:}Long-term validation of suppressed bistability}

\begin{figure}[H]
    \centering
    \includegraphics[width=0.9\linewidth]{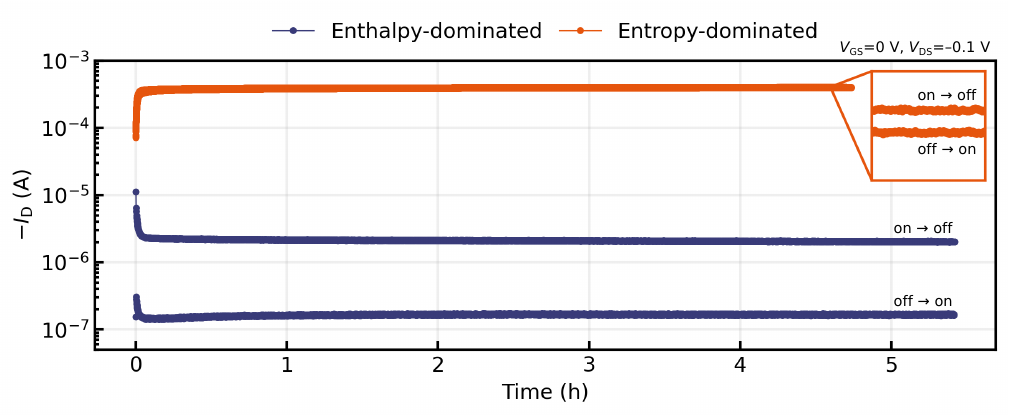} 
    \caption{\textbf{Transient state retention under enthalpic and entropic dominance.} Shielding the enthalpic interactions electrostatically diminishes the bistability, which translates to a suppressed state retention.\label{fig:S_Longterm_Compare}}
\end{figure}
\addcontentsline{toc}{subsection}{Fig.~\protect\numberline{\thefigure:}Transient state retention under enthalpic and entropic dominance}

\begin{figure}[H]
    \centering
    \includegraphics[width=\linewidth]{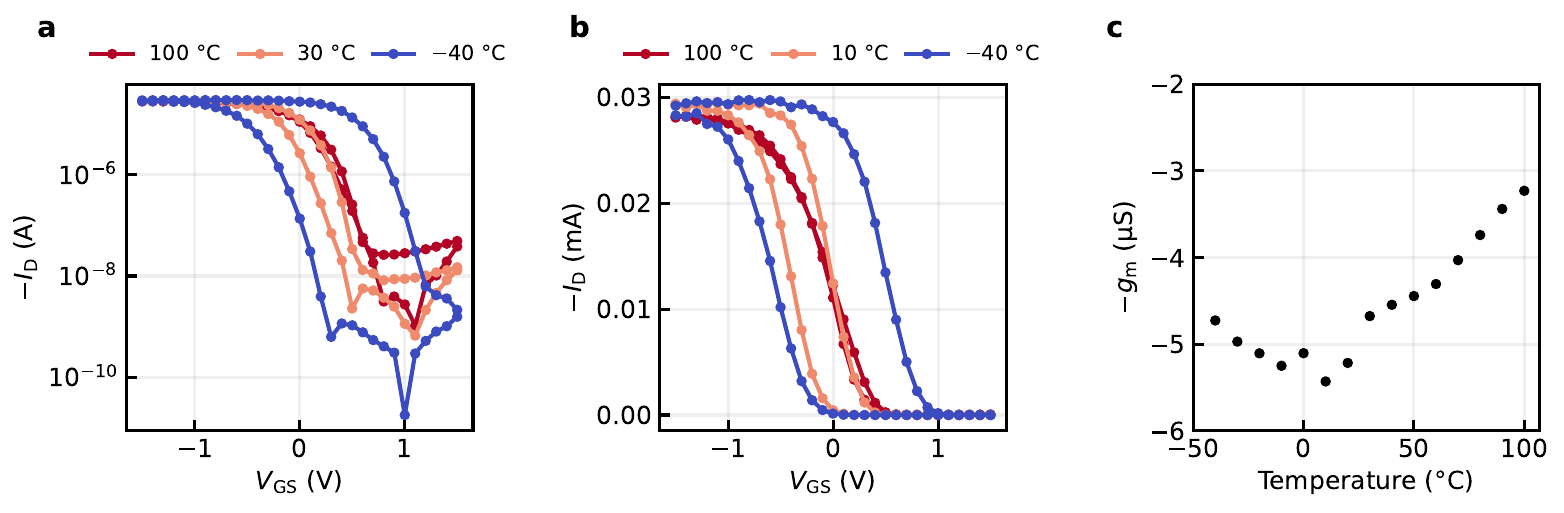} 
    \caption{\textbf{Non-monotonic dependence of the subthreshold swing on temperature.} (\textbf{a}, \textbf{b}) Temperature-dependent transfer measurements ($V_\mathrm{DS}=-0.01\,\si{\volt}$) reveal a non-monotonic progression of the subthreshold swing, going along with a steadily decreasing bistability. (\textbf{c}) Similarly, a non-monotonic dependence of the transconductance on temperature is found.\label{fig:NDC_old}}
\end{figure}
\addcontentsline{toc}{subsection}{Fig.~\protect\numberline{\thefigure:}Non-monotonic dependence of the subthreshold swing on temperature}

\begin{figure}[H]
    \centering
    \includegraphics[width=0.9\linewidth]{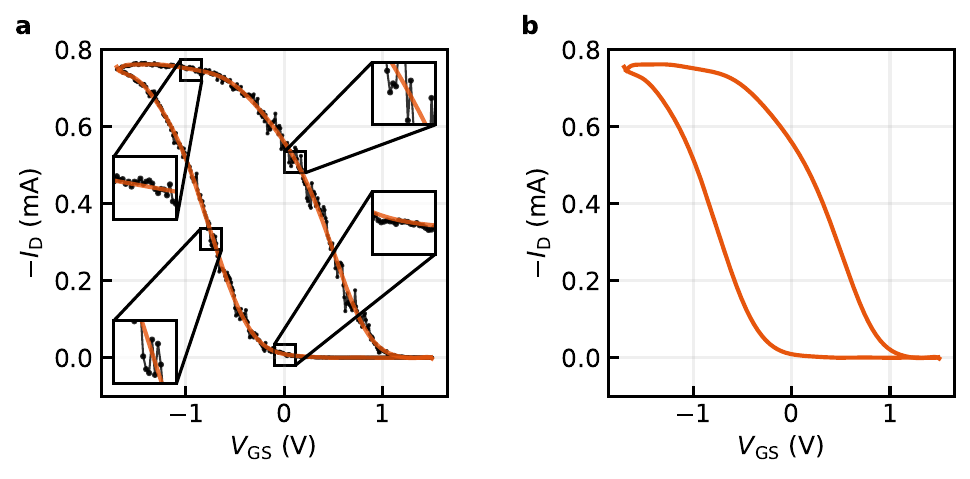} 
    \caption{\textbf{Noise extraction of Schmitt trigger.} (\textbf{a}) Analytical function overlaying the output. (\textbf{b}) The analytical function reflects the transfer curve with on- and off-state saturation, without overfitting the output noise.
    \label{fig:S_Schmitt Trigger}}
\end{figure}
\addcontentsline{toc}{subsection}{Fig.~\protect\numberline{\thefigure:}Noise extraction of Schmitt trigger}
\noindent For noise analysis, the output noise is determined against an analytic reference for practical reasons, namely a polynomial fit of order 12. Such high order was necessary to adequately represent the shape of the sweeps, including the saturation of the on- and off-states. The function shows good, balancing overlay with the raw data (Fig.\,\ref{fig:S_Schmitt Trigger}a), while at the same time does not over-fit the noise, which would indicate a disproportionately strong noise reduction (Fig.\,\ref{fig:S_Schmitt Trigger}b).

\addcontentsline{toc}{subsection}{Fig.~\protect\numberline{\protect\ref{fig:S_Circuit}}Ionic circuit of an OECT}

\addcontentsline{toc}{subsection}{Fig.~\protect\numberline{\protect\ref{fig:S_Phase portrait}}Dynamic instability}

\addcontentsline{toc}{subsection}{Fig.~\protect\numberline{\protect\ref{fig:S_ChemWork}}Chemical potential profile as a cycle process}

\addcontentsline{toc}{subsection}{Fig.~\protect\numberline{\protect\ref{fig:S_Thermography}}In-operando thermography study of a solid-state OECT}

\addcontentsline{toc}{subsection}{Fig.~\protect\numberline{\protect\ref{fig:S_SubthresholdSwing_Simulation}}Subthreshold swing of a bistable OECT (simulation)}

\addcontentsline{toc}{subsection}{Fig.~\protect\numberline{\protect\ref{fig:S_Fitting}}Extracted and fitted doping efficiencies $\alpha$}

\addcontentsline{toc}{subsection}{Fig.~\protect\numberline{\protect\ref{fig:S_DrainVoltage}}Effect of the drain voltage}

\addcontentsline{toc}{subsection}{Fig.~\protect\numberline{\protect\ref{fig:S_Proxy_Schmitt}}Dynamic response of a bistable system}
    
\addcontentsline{toc}{subsection}{Fig.~\protect\numberline{\protect\ref{fig:S_SchmittTrigger_1}}OECT oscillation}

\addcontentsline{toc}{subsection}{Fig.~\protect\numberline{\protect\ref{fig:S_SchmittTrigger_2}}First-order Butterworth filter}

\let\oldcite\cite
\renewcommand{\cite}[1]{\textsuperscript{\oldcite{#1}}}
\newcommand{\normcite}[1]{\textnormal{\oldcite{#1}}}

\clearpage
\section*{Supplementary Notes}
\addcontentsline{toc}{section}{Supplementary Notes}
\label{Note_S:Theory}

\section*{Supplementary Note 1: Theoretical Framework}
\addcontentsline{toc}{subsection}{Supplementary Note 1: Theoretical Framework}
We consider the Gibbs free energy function, defined as 
\begin{equation}
    \label{eq:S_GibbsFunction}
     G = H-TS,
\end{equation}
where $H$ is enthalpy, $T$ is temperature, and $S$ is entropy. Let the channel be composed of doping subunits, where one subunit is the smallest entity that satisfies the doping equation \begin{equation}
\label{eq:S_pedot reaction}
      \ce{
PEDOT{:}PSS + C^+ + A^- <=> PEDOT^{+}{:}PSS + C^+ + A^- + e^-,
}  
\end{equation}
where \ce{PEDOT{:}PSS} and \ce{PEDOT^{+}{:}PSS} are the initial and doped state of the OMIEC, and \ce{C^+} and \ce{A^-} are the electrolyte cat- and anions. Given this, we can regard Eq.\,\ref{eq:S_GibbsFunction} as a function of the relative share of doped units $\psi$:
\begin{equation}
    \psi = \frac{N_\mathrm{doped}}{N_\mathrm{doped}+ N_\mathrm{undoped}}= \frac{N_\mathrm{doped}}{N_\mathrm{tot}},
    \label{eq:S_def_phi}
\end{equation}
where $N_\mathrm{doped}$ and $N_\mathrm{undoped}$ are the number of doped and undoped sites and $N_\mathrm{tot}$ is the total number available. For the binary system, the Gibbs free energy function follows as 
\begin{eqnarray}
     G(\psi)& = \frac{\tilde{G}(\psi)}{N_\mathrm{tot}} = H^0(\psi) + H_\mathrm{tr}(\psi) - TS_\mathrm{tr}(\psi), \label{eq:S_Gibbs_composition}
\end{eqnarray}
with the enthalpic and entropic terms as 
 \begin{eqnarray}
       H^0(\psi) & = \psi\mu^0_{\mathrm{d}} + (1-\psi)\mu^0_{\mathrm{u}}, \label{eq:S_H0}\\
        H_\mathrm{tr}(\psi) &= \frac{Z}{2}\left(w_{dd}\psi^2 + w_{uu}(1-\psi)^2 + 2w_{du} \psi(1-\psi)\right), \quad \text{and} \label{eq:S_Hmix} \\
       S_\mathrm{tr}(\psi) & = -k_\mathrm{B} (\psi\ln(\psi) + (1-\psi)\ln(1-\psi)). \label{eq:S_Smix}
\end{eqnarray}
$H^0(\psi)$ denotes the standard enthalpy, while $H_\mathrm{tr}(\psi)$ and $S_\mathrm{tr}(\psi)$ are the enthalpy and entropy associated with the state transitions underlying Eq.\,{\ref{eq:S_pedot reaction}}. $\mu^0_{\mathrm{d}}$ and $\mu^0_{\mathrm{u}}$ are the standard chemical potentials of doped and undoped sites, $Z$ is the coordination number, and $w_{dd}$, $w_{uu}$, and $w_{du}$ express the interaction strength between doped sites, undoped sites, and between the two. $k_\mathrm{B}$ is the Boltzmann constant. The chemical potential follows as 
\begin{align} \label{eq:S_ChemPot}
\mu(\psi)  &=  \left(\frac{\partial G(\psi)}{\partial \psi}\right)_{p, T} \nonumber \\
& =  \mu^0_{\mathrm{d}} - \mu^0_{\mathrm{u}} + Z\left(w_{dd}\psi+w_{uu}(\psi-1)-w_{du}(2\psi-1)\right)+k_\mathrm{B}T\ln{\left(\frac{\psi}{1-\psi}\right)}
\end{align}
and relates to the gate-source voltage $V_\mathrm{GS}$ via the electrochemical potential as defined in Eq.\,\ref{eq:Chem_Potential}. Since the doping parameter $\psi$ translates to the drain current $I_\mathrm{D}$, the transfer curve of an OECT can be seen as a direct consequence of the underlying Gibbs free energy function. 
\nocite{bistability_simulation}
For an entropy-dominated system, $G(\psi)$ is a parabola-shaped potential with a single minimum, i.e., a single thermodynamic equilibrium state (Fig.\,\ref{fig:1}a). However, for a system of dominating enthalpy, the single equilibrium state bifurcates, given rise to a partially negative curvature in $G(\psi)$:
\begin{eqnarray}
& \left(\frac{\partial^2G(\psi)}{\partial\psi^2}\right)_{p,T} &\le0 \\
&Z (w_{dd}+w_{uu}-2w_{du})&\le\frac{k_\mathrm{B}T}{\psi(\psi-1)}  \\
\lambda = &\frac{Z (w_{dd}+w_{uu}-2w_{du}) }{k_\mathrm{B}T} \cdot \psi(\psi-1)& \geq 1 \quad \text{with} \quad \psi\in[\psi_i, 1-\psi_i],
\label{eq:S_Condition_Bistability}
\end{eqnarray}
where $\psi_i$ and $1-\psi_i$ are the inflection points of $G(\psi)$. Doping concentrations in this $\psi$-range are unstable and decompose into two coexisting equilibrium states with positive curvature in $G(\psi)$. In this situation, the chemical potential is non-monotonic, having a range of inverted slope between its local extrema at $\psi_i$ and $1-\psi_i$ that cannot contribute to the static transfer curve of an OECT. The quantity $\lambda$ can in this sense be interpreted to express the degree of bistability present in the system, as it sets enthalpic and entropic contributions in relation to one another. We provide an interactive simulation tool under Ref.\,\citen{bistability_simulation} to illustrate these relationships.

Given that a single doping unit will involve multiple individual components (e.g., PEDOT units, PSS units, ions), we approximate the standard chemical potentials with $\mu^0_{\mathrm{d}}\approx\mu^0_{\mathrm{u}}$, from which an equilibrium at $\psi=0.5$ follows for the ideal gas scenario. Apart from that, it is worth noting that the approach we take here is based on the assumption of a lattice structure, where $Z$ is the according coordination number. Obviously, the OMIEC's microstructure does not correspond to a period lattice but is instead of a much more disordered nature, which translates to the interlacing of doping units. Accordingly, $Z$ appears as an unknown quantity that we cannot explicitly separate from the interaction parameters, but only consider the product with
\begin{equation}
    h_i = Z \cdot w_i \quad \forall \quad i \in \{dd, uu, du\},
\end{equation}
where for $Z$ we expect values of typical scale for three-dimensional systems, i.e., 2 to $\sim8$.

\clearpage
\section*{Supplementary Note 2: Dynamic Instability}
\addcontentsline{toc}{subsection}{Supplementary Note 2: Dynamic Instability}
\label{Note_S:Dynamics_Instability}

\begin{figure}[H]
    \centering
    \includegraphics[width=0.35\linewidth]{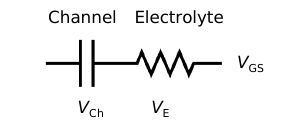} 
    \caption{\textbf{Ionic circuit of an OECT.}
    \label{fig:S_Circuit}}
\end{figure}

With $\psi$ describing the relative proportion of doped units in the channel, the associated charge follows as
\begin{equation}
    \varphi = \psi\cdot e \cdot N_\mathrm{tot}.
\end{equation}
As the chemical potential $\mu(\varphi)$ is related to $V_\mathrm{GS}$ through Eq.\,\ref{eq:Chem_Potential}, the inverse derivative of $\mu(\varphi)$ has the character of a capacitance,
\begin{equation}
    \frac{\partial\varphi}{\partial\mu(\varphi)}  =C(\varphi).
    \label{eq:S_capacitance_proxy}
\end{equation}
From this, we can express the time dependence of $\varphi$ through
\begin{equation}
\label{eq:S_phi/dt}
    \frac{\partial\varphi}{\partial t} = C(\varphi)\frac{\partial\mu(\varphi)}{\partial t}.
\end{equation}
 To now understand the dynamic behavior of the channel during a transfer scan, we take the approach of Bernards et al.\autocite{bernards2007steady} and describe the system as a series connection of capacitor and resistor, where we assume a non-polarizable gate electrode (e.g., Ag/AgCl) for the sake of simplicity (Fig.\,\ref{fig:S_Circuit}). Given this system, $\frac{\partial\varphi}{\partial t}$ (Eq.\,\ref{eq:S_phi/dt}) corresponds to the gate current $I_\mathrm{G}$, which itself is determined by the effective voltage through the electrolyte $V_\mathrm{E}$:
\begin{align}
    I_\mathrm{G}    & \leftrightarrow \frac{\partial\varphi}{\partial t} \\
                    & = \frac{1}{R_\mathrm{E}} V_\mathrm{E} \\ 
                    & = \frac{1}{R_\mathrm{E}} (V_\mathrm{GS}-V_\mathrm{Ch}).
\end{align}      
Obviously, the dynamic behavior of $I_\mathrm{G}$ during a $V_\mathrm{GS}$-sweep also determines the charge in the channel and therefore, $I_\mathrm{D}$. When further considering that the channel potential is predominantly determined by the chemical potential $\mu(\varphi)$, the OECT dynamics can be described through the first-order differential equation
\begin{equation}
    I_\mathrm{D}     \leftrightarrow \frac{\partial\varphi}{\partial t}
                   = \frac{1}{R_\mathrm{E}} (V_\mathrm{GS}-\mu(\varphi)).
\end{equation}

\begin{figure}[t!]
    \centering
    \includegraphics[width=\linewidth]{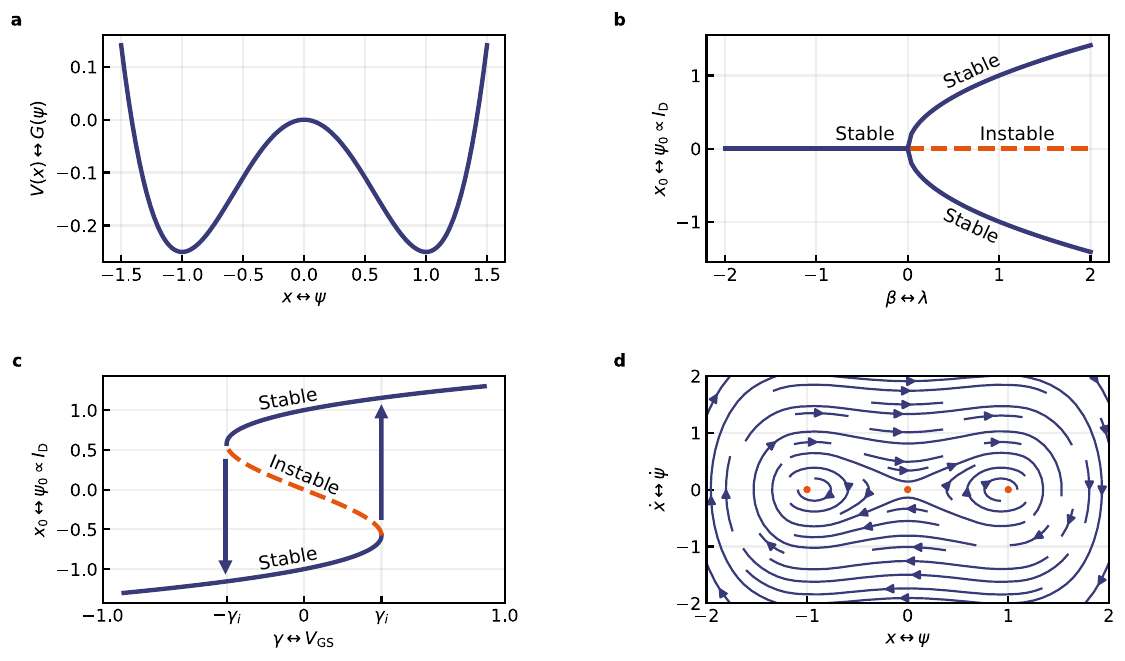} 
    \caption{\textbf{Dynamic instability.} (\textbf{a}) $G(\psi)$ is approached by a fourth-order polynomial $V(x)$, according to Eq.\,\ref{eq:S_Taylor} ($\alpha=0.25, \beta=0.5$). (\textbf{b}) Increasing $\beta$ effects a pitchfork bifurcation of the equilibrium state for $\beta>0$. (\textbf{c}) For an external force $\gamma$, stable and instable equilibrium regimes occur, yielding a hysteresis. (\textbf{d}) Phase portrait visualizing the dynamics of the bistable system, showing an unstable fixed point in between two stable fixed points.
    \label{fig:S_Phase portrait}}
\end{figure}

\noindent For a given bistability (Eq.\,\ref{eq:Condition_Bistability}), one can approximate the bistable potential function as a polynomial of order 4, namely
\begin{equation}
\label{eq:S_Taylor}
    G(\psi) \leftrightarrow  V(x) = \alpha x^4 - \beta x^2, \quad \text{where} \quad \alpha, \beta>0.
\end{equation}
Given such a potential (Fig.\,\ref{fig:S_Phase portrait}a), one can understand the dynamic behavior of the bistable system by comparing it to a damped, non-linear oscillator. With including a damping proportional to the velocity, it follows from Newton's second law that 
\begin{eqnarray}
\label{eq:S_Newton}
\ddot{x} & = -\frac{\partial V(x)}{\partial x}-\eta\dot{x}, \quad \text{which for strong damping leads to} \\
\label{eq:S_Newton_FirstOrder}
\dot{x} & = -\frac{\partial V(x)}{\partial x} = -4\alpha x^3+2\beta x
\end{eqnarray}
on the new time axis $t\rightarrow t\eta^{-1}$. Eq.\,\ref{eq:S_Newton_FirstOrder} is equal to the force driving the system into its equilibrium states, i.e., the chemical potential $\mu(\psi)$, and as such, is a gradient system. This implies asymptotically stable fixed points of $V(x)$ at $x_0=\pm\sqrt{\beta(2\alpha)^{-1}}$ and an instable point at $x_0=0$, leading to bifurcation depending on $\beta$ (Fig.\,\ref{fig:S_Phase portrait}b). With the two stable points being separated by an instable point, bistable operation follows. When the system is deflected from equilibrium by a constant, external force $\gamma$, Eq.\,\ref{eq:S_Newton_FirstOrder} turns to 
\begin{eqnarray}
\dot{x} & = -\frac{\partial V(x)}{\partial x} = -4\alpha x^3+2\beta x+\gamma,
\end{eqnarray}
leading to bistability in the interval of $\gamma \in (-\gamma_i, \gamma_i)$ with $\gamma_i^2 = 8\beta^3(27\alpha)^{-1}$ (Fig.\,\ref{fig:S_Phase portrait}c). 
The parameter $\gamma$ will push the system always on a stable path, until either $\gamma_i$ (from the bottom) or $-\gamma_i$ (from the top) is reached, upon which the transition follows on a very short time scale. This dynamic behavior is visualised by the phase portrait in Fig.\,\ref{fig:S_Phase portrait}d, showing the two stable fixed points sideways to the unstable fixed point. We further expand on this reasoning in \hyperref[Note_S:Schmitt_Trigger]{Supplementary Note 7} to study the dynamic response of the system to a periodic bias.

Transferring this understanding to the OECT system is now straightforward. The double-well potential $V(x)$ approximates the Gibbs free energy function $G(\psi)$, having its equilibrium state at $\psi_0$ for the monostable (entropy-dominated) system and at $\psi_0$ and $1-\psi_0$ for the bistable (enthalpy-dominated) case. The bifurcation parameter $\beta$ relates to the balance between these forces, entropy and enthalpy, and thus to $\lambda$ of Eq.\,\ref{eq:Condition_Bistability}, while the  external force $\gamma$ corresponds to the gate voltage $V_\mathrm{GS}$. It follows that any system obeying the first-order ordinary differential equation derived above will show bistable operation, if the underlying potential function features the topology of Eq.\,\ref{eq:S_Taylor} (double-well potential). It further follows from this argumentation that a depletion mode device must necessarily show non-volatile hysteresis in the loop direction of Fig.\,\ref{fig:1}c. That is, a low-resistance state followed by a high-resistance state, when switching from on to off.

\clearpage
\section*{Supplementary Note 3: Maxwell Construction}
\addcontentsline{toc}{subsection}{Supplementary Note 3: Maxwell Construction}
\label{Note_S:Maxwell}
\begin{figure}[h!]
    \centering
    \includegraphics[width=0.6\linewidth]{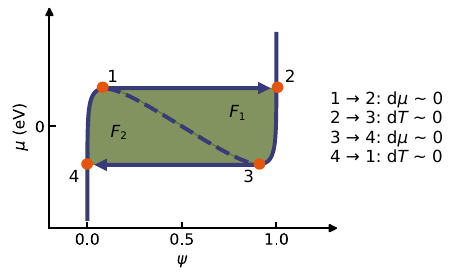} 
    \caption{\textbf{Chemical potential profile as a cycle process.} \label{fig:S_ChemWork}}
\end{figure}

\begin{figure}[H]
    \centering
    \includegraphics[width=\linewidth]{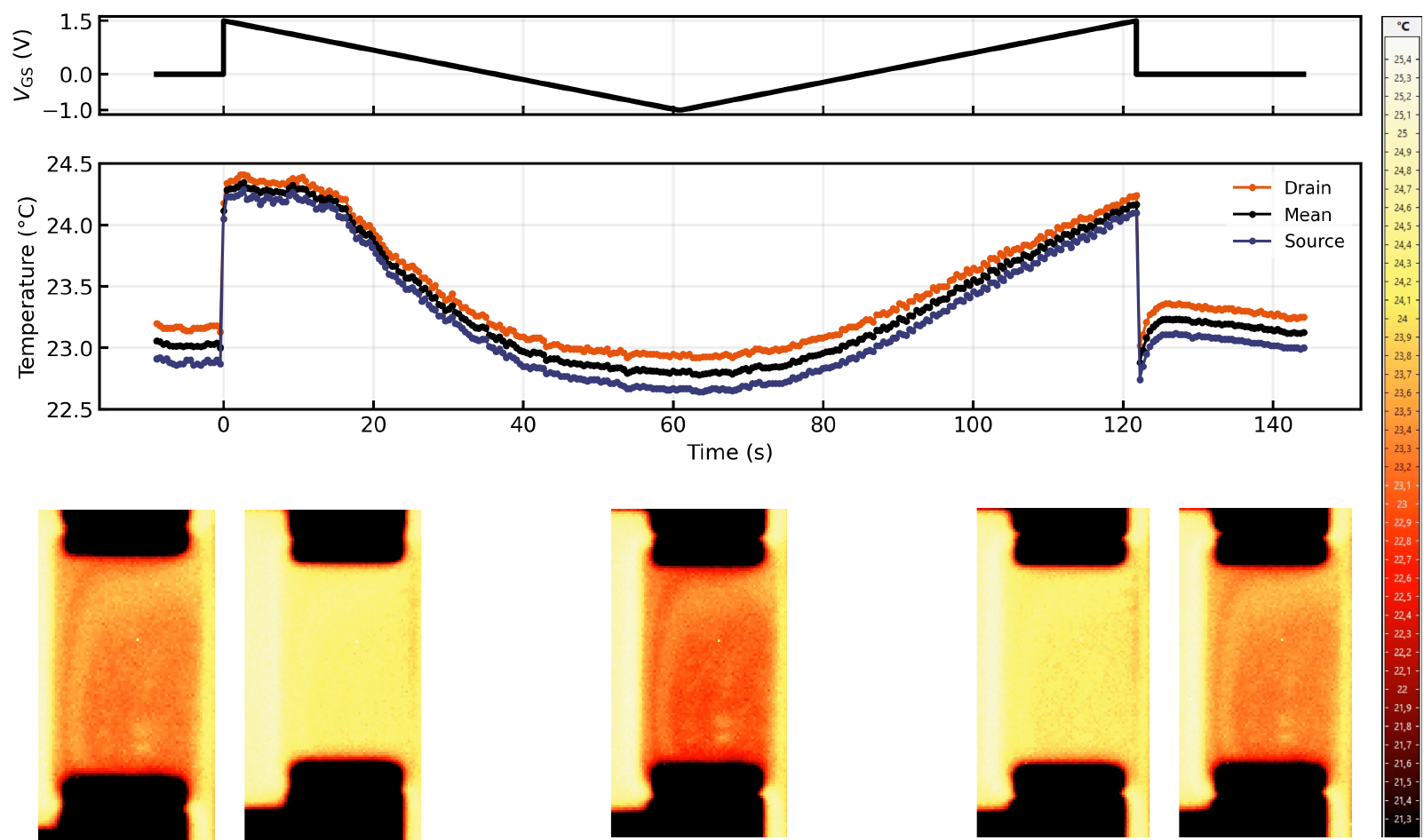} 
    \caption{\textbf{In-operando thermography study of a solid-state OECT.}  Heat consumption is observed when switching the OECT from off to on ($V_\mathrm{GS}=1.5$ to $-1.0\,\si{\volt}$, $V_\mathrm{DS}=-0.1\,\si{\volt}$). For the time being, differences between source and drain electrodes are omitted and the temperature is averaged over the channel. We attribute the temperature difference, which in the on-state is $\sim0.3\,\si{\kelvin}$, to a shift of the doping process towards the source electrode, at which charge injection takes place.\label{fig:S_Thermography}}
\end{figure}

\noindent We show that the profile of $\mu(\psi)$ can be considered as a cycle process\autocite{borgnakke2022fundamentals}. As laid out in the main text and in \hyperref[Note_S:Theory]{Supplementary Note 1}, we assume similar standard chemical potential for doped and undoped units, which allows us to approximate the idealized chemical potential point-symmetric around $(\psi, \mu(\psi)) = (0.5, 0\,\text{eV})$. The first law of thermodynamics dictates that
\begin{eqnarray}
    \label{eq:S_1stLawTD}
    \mathrm{d}U & = \delta Q + \delta W,
\end{eqnarray}
with the internal energy $U$ and $\delta Q$ and $\delta W$ the heat exchanged and work performed. The second law of thermodynamics demands for a reversible process that
\begin{eqnarray}
    \label{eq:S_2ndLawTD}
    \delta Q & = T\mathrm{d}S,
\end{eqnarray}
while the performed chemical work corresponds to
\begin{eqnarray}
    \label{eq:S_ChemWork}
    \delta W & = \int \mu(\psi) \mathrm{d}\psi, \quad \text{yielding} \\
    \mathrm{d}U & = T\mathrm{d}S +\int \mu(\psi) \mathrm{d}\psi \quad \text{and} \\
    \mathrm{d}(U-TS) & = -S\mathrm{d}T+\mu(\psi)\mathrm{d}\psi.
\end{eqnarray}
With the Helmholtz free energy defined as 
\begin{eqnarray}
    \label{eq:s_DefFreeEnergy_1}
    F & = U-TS, \quad \text{it follows that} \\
    \mathrm{d}F & = -S\mathrm{d}T+\mu(\psi)\mathrm{d}\psi, \label{eq:s_DefFreeEnergy_2}
\end{eqnarray}
where the first term expresses the exchanged heat and the second the performance of chemical work, corresponding to $F_1$ and $F_2$ in Fig.\,\ref{fig:S_ChemWork}:
\begin{eqnarray}
    F_1 & = F_2 \\
 \int_1^2\mu_{1\rightarrow2}(\psi) \ \mathrm{d}\psi &= \int_3^4\mu_{3\rightarrow4}(\psi) \ \mathrm{d}\psi.
\end{eqnarray}
This reasoning is similar to the concept of Maxwell constructions in the context of phase transitions of non-ideal gases, where the work done during expansion and compression must be equal. Here, however, we refer to the chemical work. From Eq.\,\ref{eq:s_DefFreeEnergy_2} follows that the performance of such work must necessarily be associated with the consumption/release of heat, which we validate experimentally in Fig.\,\ref{fig:3}g and \ref{fig:S_Thermography}. 

\clearpage
\section*{Supplementary Note 4: Subthreshold Swing}
\addcontentsline{toc}{subsection}{Supplementary Note 4: Subthreshold Swing}
\label{Note_S:SubthresholdSwing}

\begin{figure}[H]
    \centering
    \includegraphics[width=\linewidth]{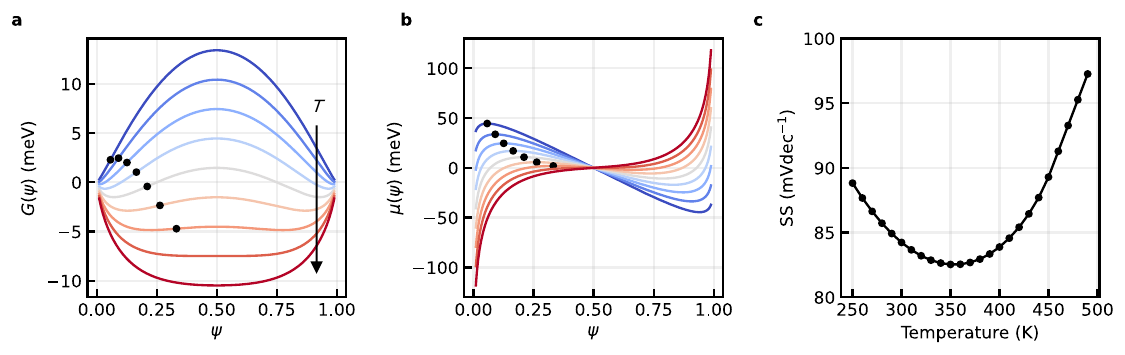} 
    \caption{\textbf{Subthreshold swing of a bistable OECT (simulation).} With a bistable Gibbs free energy function (\textbf{a}), the chemical potential (\textbf{b}) is non-monotonic. During the transition between the on- and off-state, the total electrochemical potential is constant, implying that chemical potential and electrostatic potential balance (Eq.\,\ref{eq:S_ChemPot_Balance}). This effect decreases with rising temperature, causing a non-monotonic progression of the subthreshold swing with temperature. We express this relationship through Eq.\,\ref{eq:S_OECTSS}, which is modelled in (\textbf{c}). 
  \label{fig:S_SubthresholdSwing_Simulation}}
\end{figure}
As laid out before, for a present bistability, we can assume a Maxwell construction for $\mu(\psi)$ in the regime $\psi\in[\psi_i, 1-\psi_i]$, where the total electrochemical potential is constant during each transition (Fig.\,\ref{fig:S_ChemWork}). Following Eq.\,\ref{eq:Chem_Potential}, this is the result from a balance between the chemical potential resulting from $G(\psi)$ and the electrostatic potential from the gate voltage $V_\mathrm{GS}$:
\begin{equation}
    \bar{\mu}=\frac{\partial G(\psi)}{\partial\psi} +fe(V_\mathrm{GS}-V_\mathrm{Ch}) = \mathrm{const.}
    \label{eq:S_ChemPot_Balance}
\end{equation}
This understanding can be transferred to the subthreshold swing (SS), which defines the minimum gate voltage required to change the drain current by one order of magnitude in the subthreshold region. In the ideal case, the subthreshold swing is a purely diffusion-controlled quantity, being inversely proportional to the thermal voltage $\frac{k_\mathrm{B}T}{e}$, which defines the typically expected increase with temperature\autocite{sze2021physics}: As the temperature increases, a progressively larger gate potential is required to change the drain current by one order of magnitude, since diffusion deprives charge carriers from their effective modulation. For the bistable OECT system, however, another effect is to be considered: With the non-monotonic chemical potential compensating the electrostatic gate potential, the subthreshold swing is subject to a second influence, which ought to be stronger at lower temperatures. Thus, the subthreshold swing is expected to obey 
\begin{equation}
    \label{eq:S_OECTSS}
    \mathrm{SS}(T)  \approx \frac{\ln(10)}{e} \left( \left(\left. \frac{\partial G(\psi, T)}{\partial \psi} \right|_{\psi =\psi_i}\right)_{p,T} +k_\mathrm{B}T \right),
\end{equation}
where we can assume a doping efficiency of $f\approx1$, motivated by the low state occupancy in the subthreshold region. The chemical potential is evaluated at its local maximum $\psi_i$, which is due to the fact that the subthreshold swing for the PEDOT:PSS-based depletion mode OECT is considered for its transition from on to off (Fig.\,\ref{fig:S_SubthresholdSwing_Simulation}a,b). $\psi_i$ in this case is the first thermodynamically stable point, where at the same time, chemical and electrostatic potential are just balanced. Eq.\,\ref{eq:S_OECTSS} is modelled in Fig.\,\ref{fig:S_SubthresholdSwing_Simulation}c, showing the non-monotonic dependence on temperature: As long as enthalpy is dominant and bistability is present, the subthreshold swing decreases with rising temperature. From the point on where enthalpic and entropic contributions are balanced, however, the subthreshold swing increases with $k_\mathrm{B}T$. 

\clearpage
\section*{Supplementary Note 5: Fitting of Gibbs Free Energy}
\addcontentsline{toc}{subsection}{Supplementary Note 5: Fitting of Gibbs Free Energy}
\label{Note_S:G_Fitting}

Calculating and fitting the experimental Gibbs free energy data was carried out on the basis of the procedures of Ref.\,\citen{cucchi2022thermodynamics}. However, certain refinements were necessary to account for the bidirectional device operation and the resulting bistability underlying this work. In particular, we model the Gibbs free energy profiles of the data sets of Fig.\,\ref{fig:3}a (entropy-controlled) and b (enthalpy-controlled). $G(\psi)$ can be extracted from the transfer curve measurements via integration. Especially for the low-temperature regime, however, the given electrochemical window does not allow to capture both the on- and the off-state symmetrically, which may hinder appropriate data integration. To account for this, we perform an analytic continuation of the data in the on-state, by fitting the two branches with a generalized logistic function of the form
\begin{equation}
    f(x) = \frac{I_\mathrm{on}}{(1 + e^{-b \cdot (x - c)})^v} + I_\mathrm{off},
\end{equation}
where $x$ refers to $V_\mathrm{GS}$ and the fitting parameters $b$, $c$, and $v$ account for the fitting functions slope, midpoint, and sharpness. The normalized data sets are then extended in an equidistant manner to achieve symmetry around the midpoints at $\psi=0.5$. An example is shown in Fig.\,\ref{fig:3}c, where the dark blue data set is continued with the light blue data points. Obviously, this approach preempts any processes potentially occurring at more negative gate voltages that are not covered by the experimental data, like the formation of bipolarons. However, since such processes are not described by our model anyway at this point, the analytic continuation of the on-state as carried out here represents a valid approximation. Given this, the reasoning of \hyperref[Note_S:Maxwell]{Supplementary Note 3} is applied by shifting the data sets to have partial integrals of equal size around $\mu(\psi)=0\,\si{\electronvolt}$ (Fig.\,\ref{fig:3}c). Note that this approach is underpinned by the approximately equal heat exchange for the two sweeps in the thermography study (Fig.\,\ref{fig:3}g and \ref{fig:S_Thermography}). Data integration then yields the experimental Gibbs free energy profiles shown in Fig.\,\ref{fig:3}d and e with hollow points. 

For fitting, the model equations (Eq.\,\ref{eq:Gibbs_Intro}-\ref{eq:S_mix}) were subjected to polynomial decomposition to then fit the equation
\begin{equation}
    g(\psi) = \frac{1}{\alpha} \left(k_\mathrm{B}T(\psi\ln(\psi)+(1-\psi)\ln(1-\psi))  + p\psi^2 + q\psi + r\right)
\label{eq:S_GFit}
\end{equation}
with a weighted Levenberg-Marquardt algorithm, where $\alpha$ accounts for the fudge factor of Eq.\,\ref{eq:Chem_Potential}. Data points were weighted according to 
\begin{equation}
    w_i=\left(\frac{d\mu_i}{d\psi_i}\right)^2,
\end{equation}
before being normalized. From Eq.\,\ref{eq:S_GFit}, the $h$-parameters can be extracted by considering that the Gibbs free energy function can be fully described by three points
\begin{align}
    g_0 & = G\left(\psi = 0\right) = \mu^0_{u} +\frac{1}{2}h_{uu}, \\
    g_1 & = G\left(\psi = 1\right) = \mu^0_{d} + \frac{1}{2}h_{dd}, \quad \text{and,}\\
    g_2 & = G\left(\psi = \frac{1}{2}\right) = (h_{dd}+h_{uu}-2h_{du}) = -8\left[G\left(\psi = \frac{1}{2}\right)-\frac{1}{2}\left(g_0+g_1\right)+\ln(2)k_\mathrm{B}T\right],
\end{align}
from which $h_{dd}$, $h_{uu}$, and $h_{du}$ follow. The results are summarized in Table\,\ref{tab:S_Fitting_h} and \ref{tab:S_Fitting_alpha}, together with the degree of bistability $\lambda$, as calculated from Eq.\,\ref{eq:Condition_Bistability} for $\psi = 0.5$.
\begin{table}[H]
\caption{\textbf{Fitting parameters and extracted interaction parameters $\bm{h}$.} The data set-crossing $h$-parameters are obtained by a global fitting approach. Accurate $\lambda$ values are given in Table\,\ref{tab:S_Fitting_alpha}.}
\centering
\begin{tabular}{r c c c c}
\hline
System & $h_{dd}\,(\si{\milli\electronvolt})$ & $h_{uu}\,(\si{\milli\electronvolt})$ & $h_{du}\,(\si{\milli\electronvolt})$ & $\lambda$ \\
\hline
$TS_\mathrm{tr}$ (Fig.\,\ref{fig:3}d) & $3.124\pm0.023$ &$4.036\pm0.034$ & $61.625\pm0.142$ & $>1$ \\
$H_\mathrm{tr}$ high (Fig.\,\ref{fig:3}e) & $2.195\pm0.013$ &$6.560\pm0.055$ &  $68.853\pm0.070$ & $>1$\\
$H_\mathrm{tr}$ low (Fig.\,\ref{fig:3}e) & $2.194\pm0.025$ &$2.391\pm0.038$ &  $42.336\pm0.512$ & $<1$\\
\hline
\end{tabular}
\label{tab:S_Fitting_h}
\end{table}

 We note that visually, the entropy-controlled data sets (Fig.\,\ref{fig:3}d) have poorer agreement with the fits than the enthalpy-controlled ones (Fig.\,\ref{fig:3}e). This is due to the fact that, for the time being, the model presented herein does not assume any temperature-dependence of the $h$-parameters, for which the only free fitting parameter for each individual data set is $\alpha$. The $h$-parameters, on the other hand, must cover all data sets concurrently. The fitting of the enthalpy-controlled data sets is not subject to such constraints, which allows for the better match.

\clearpage
\section*{Supplementary Note 6: Doping Efficiency}
\addcontentsline{toc}{subsection}{Supplementary Note 6: Doping Efficiency}
\label{Note_S:doping_efficiency}

\begin{table}[H]
\caption{\textbf{Extracted doping efficiencies $\bm{\alpha}$ and degrees of bistability $\bm{\lambda}$.} $\lambda$ is calculated following Eq.\,\ref{eq:Condition_Bistability} with $\psi = 0.5$.}
\centering
\begin{tabular}{r c c c }
\hline
System & Temperature & $\alpha\,(\si{\electronvolt\per\volt})$ & $\lambda$ \\
\hline
$TS_\mathrm{tr}$ (Fig.\,\ref{fig:3}d) & 263\,K & $0.0078\pm0.0001$ & 1.281 \\ 
& 268\,K & $0.0161\pm0.0001$ & 1.257 \\
& 273\,K & $0.0179\pm0.0001$ & 1.234 \\
& 278\,K & $0.0138\pm0.0001$ & 1.212 \\
& 283\,K & $0.0132\pm0.0001$ & 1.191 \\
& 288\,K & $0.0177\pm0.0001$ & 1.170 \\
& 293\,K & $0.0211\pm0.0001$ & 1.150 \\
& 298\,K & $0.0264\pm0.0002$ & 1.131 \\
& 303\,K & $0.0320\pm0.0003$ & 1.112 \\
$H_\mathrm{tr}$ high (Fig.\,\ref{fig:3}e) & 293\,K & $0.0223\pm0.0002$ & 1.277 \\
$H_\mathrm{tr}$ low (Fig.\,\ref{fig:3}e) & 293\,K & $0.0506\pm0.0005$ &  0.793\\
\hline
\end{tabular}
\label{tab:S_Fitting_alpha}
\end{table}

\begin{figure}[t!]
    \centering
    \includegraphics[width=0.5\linewidth]{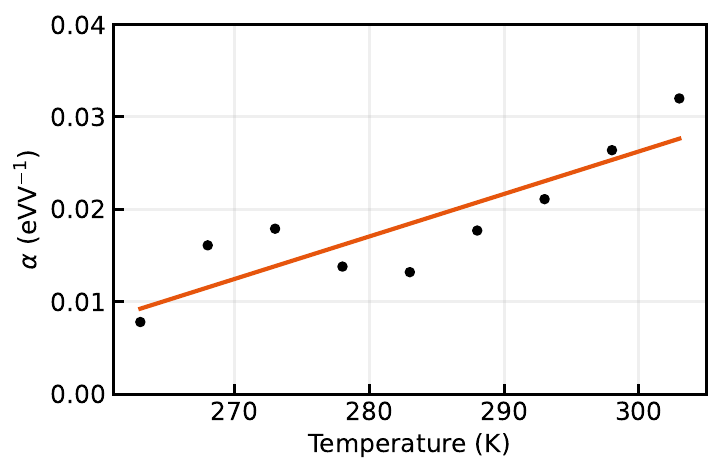} 
    \caption{\textbf{Extracted and fitted doping efficiencies $\bm{\alpha}$.} The doping efficiencies $\alpha$ from Table\,\ref{tab:S_Fitting_alpha} are fitted with $\alpha(T)=k_\mathrm{B}Te^{-1} \cdot b + c$, according to Eq.\,\ref{eq:S_alphaT_1} and \ref{eq:S_alphaT_2} ($b=8.55\cdot10^{-19}, c=-0.11$).  
  \label{fig:S_Fitting}}
\end{figure}

As can already be inferred from the order of magnitude of the fundamental model equations, there is a scaling involved between theoretical and experimental findings. This is reflected by the $\alpha$-parameters of Table\,\ref{tab:S_Fitting_alpha}. While these appear rather low at first glance, they can be readily understood by following reasoning:

Let the OECT gating be considered as the capacitive addition of charges. The charge carrier density $n$ then changes as
\begin{equation}
    n = n_0 + \frac{C^*\cdot V}{e}
\label{eq:S_charges_capacitive}
\end{equation}
with the applied voltage $V$, where $n_0$ is the initial charge carrier concentration and $C^*$ is the volumetric capacitance. Similarly, $n$ is changed by changing the chemical potential $\mu$, which can be expressed through the linear approximation of
\begin{equation}
    n = n_0 + \left.\frac{dn}{d\mu}\right|_{\mu=\mu_0} \cdot (\mu_0 + \Delta\mu) + \mathcal{O}(\Delta\mu^n),
\label{eq:S_charges_linear_approx}
\end{equation}
where $\mu_0$ is the initial chemical potential. The state occupation can be approximated by a Boltzmann distribution with
\begin{equation}
    n = N_\mathrm{eff}\cdot \exp{\left(\frac{E-\mu}{k_\mathrm{B}T}\right)},
\end{equation}
where $N_\mathrm{eff}$ is the effective density of states in the valence band, $E$ is the according energy, and $\mu$ is the Fermi level (i.e., chemical potential). Note that this expression refers to (electron) holes. With Eq.\,\ref{eq:S_charges_linear_approx} it follows
\begin{align}
    n &= \underbrace{N_\mathrm{eff}\cdot \exp{\left(\frac{E-\mu_0}{k_\mathrm{B}T}\right)}}_{n_0} - \frac{1}{k_\mathrm{B}T}\cdot\underbrace{N_\mathrm{eff}\cdot \exp{\left(\frac{E-\mu_0}{k_\mathrm{B}T}\right)}}_{n_0}\cdot\left(\mu_0 + \Delta\mu\right) \\
    &= n_0 \left(1-\frac{\mu_0+\Delta\mu}{k_\mathrm{B}T}\right).
\end{align}
With Eq.\,\ref{eq:S_charges_capacitive}, the change in chemical potential can then be linked to the applied voltage:
\begin{align}
    \mu_0 + \Delta\mu &= -\frac{k_\mathrm{B}T}{e} \cdot\frac{C^*}{n_0}\cdot V \label{eq:S_alphaT_1} \\
    &= -\alpha \cdot V. \label{eq:S_alphaT_2}
\end{align}
Assuming a volumetric capacitance of $C^*=40 \,\si{\farad\per\cubic\centi\metre}$, a charge carrier concentration of $n_0=1\cdot10^{21}\,\si{\per\cubic\centi\metre}$, and $k_\mathrm{B}T = 25.7\,\si{\milli\electronvolt}$ ($T=298\,\si{\kelvin}$), the scaling factor results as $0.006\,\si{\electronvolt\per\volt}$, close to the estimated factors of Table\,\ref{tab:S_Fitting_alpha}. Note that this is only an approximate estimate to indicate the order of magnitude of the scaling factor. In fact, it is not unlikely that the actual volumetric capacitance is noticeably larger, as has been shown for the treatment of PEDOT:PSS with ionic liquids\autocite{wu2019ionic}, which would result in values even closer to the ones found. The estimated relation also reflects the temperature-dependence of $\alpha$ we see in the data (Fig.\,\ref{fig:S_Fitting}), where the fitted factor of approximately $10^{-20}$ is sufficiently matching the ratio of $C^*$ to $n_0$. Not surprisingly, the data from Table\,\ref{tab:S_Fitting_alpha} finally also shows that the material composition has a notable influence on the doping efficiency ($H_\mathrm{tr}$ high vs. $H_\mathrm{tr}$ low). Not least, we infer a connection here to the capacitance described in Eq.\,\ref{eq:S_capacitance_proxy}. 

Finally, it is worth pointing out that the above calculation is valid within the Boltzmann approximation, assuming an effective density of states model for the valence band. This applies to the on-state of the OECT, where the number of holes present in the valence band is lower than the effective density of states. However, this approximation does not hold in the subthreshold region, where the presence of intra-gap states (i.e., tail states) are causing an exponential rise in current. Here, a significantly higher doping efficiency must be assumed, which is reflected experimentally by the low subthreshold slope (e.g., Fig.\,\ref{fig:3}h).

\clearpage
\section*{Supplementary Note 7: Effect of the Drain Voltage}
\addcontentsline{toc}{subsection}{Supplementary Note 7: Effect of the Drain Voltage}
\label{Note_S:DrainVoltage}

\begin{figure}[H]
    \centering
    \includegraphics[width=\linewidth]{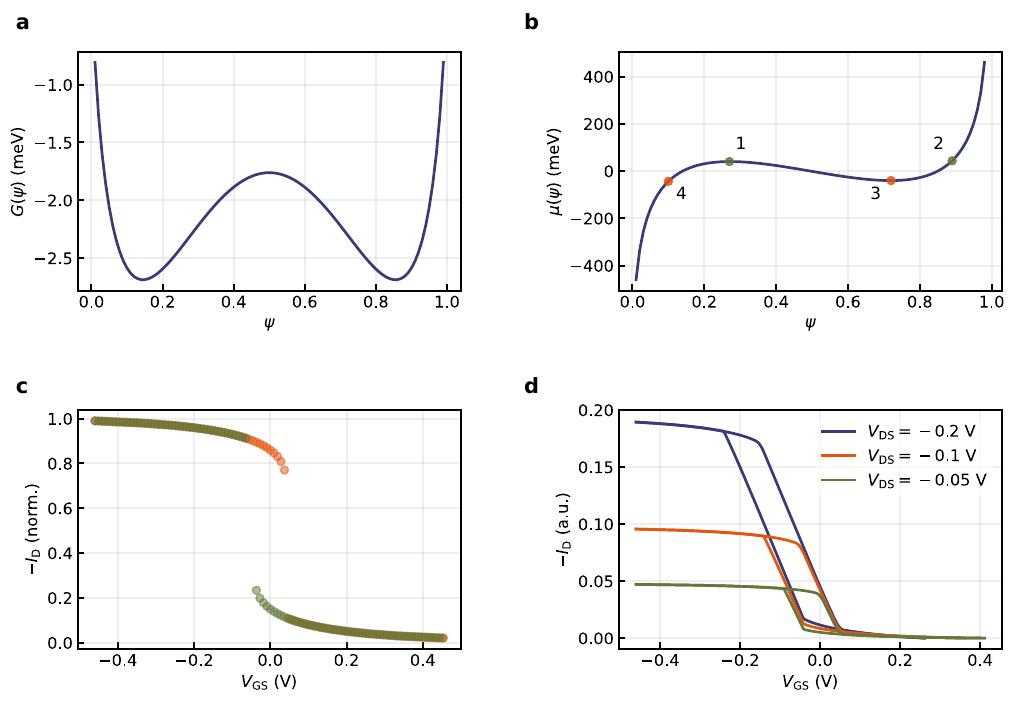} 
    \caption{\textbf{Effect of the drain voltage.} Given a bistability (\textbf{a}), the chemical potential is non-monotonic, with the four indicated points in (\textbf{b}) defining the transition points across the instable range of inverted slope. (\textbf{c}) During a transfer sweep, the physically stable points result in two branches for each sweep. Note that the orange path (dedoping) is partially covered by the green path (doping). (\textbf{d}) As the drain current is also a function of the drain voltage, the transition within each sweep deviates from a perpendicular leap.\label{fig:S_DrainVoltage}}
\end{figure}

\noindent The Gibbs free energy function of the bistable system results in a non-monotonic chemical potential profile, as derived above (Fig.\,\ref{fig:S_DrainVoltage}a,b). Since the range of inverted slope is instable, there are four points to consider for understanding the shape of the experimental transfer curve. During doping ($\psi = 0 \rightarrow \psi = 1$), the profile of the chemical potential is followed until point 1 in Fig.\,\ref{fig:S_DrainVoltage}b is reached. With $V_\mathrm{GS}$ being continuously raised, the next possible point on this curve is point 2, having the same chemical potential $\mu(\psi)$ at higher $\psi$. During dedoping ($\psi = 1 \rightarrow \psi = 0$), the same applies to points 3 and 4, respectively. This gives rise to two stable branches for each sweep, as shown in Fig.\,\ref{fig:S_DrainVoltage}c. Note that the orange track (dedoping) is largely covered by the green track (doping). To understand the transition within each sweep, one needs to consider that the drain current $I_\mathrm{D}$ results from the integral of $\psi(\mu)$ from $V_\mathrm{GS}$ to $(V_\mathrm{GS}-V_\mathrm{DS})$, as discussed in Ref.\,\citen{cucchi2022thermodynamics}. The slope of the transition is accordingly affected by $V_\mathrm{DS}$, leading to a deviation from the perpendicular leap that one might expect from other bistable systems\autocite{dreyer2010thermodynamic, koulakov2002model, vela2014key} (Fig.\,\ref{fig:S_DrainVoltage}d).

\clearpage

\section*{Supplementary Note 8: Schmitt Trigger}
\addcontentsline{toc}{subsection}{Supplementary Note 8: Schmitt Trigger}
\label{Note_S:Schmitt_Trigger}

\begin{figure}[H]
    \centering
    \includegraphics[width=\linewidth]{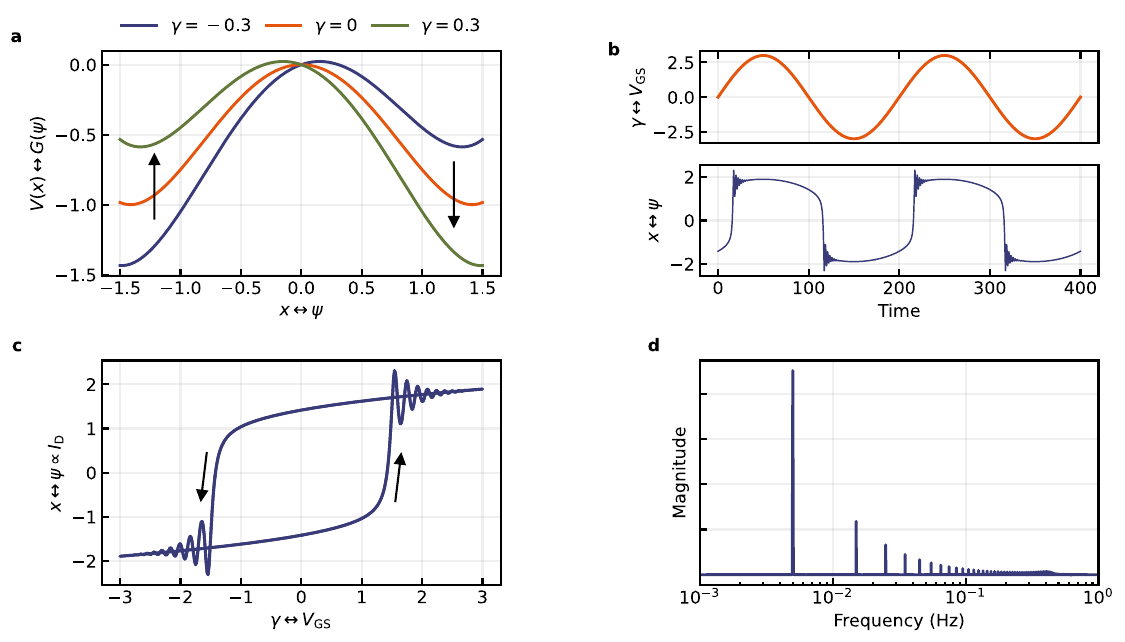} 
    \caption{\textbf{Dynamic response of a bistable system.} (\textbf{a}) $V(x)$ with $\alpha=0.25, \beta=0.5$ under an external bias $\gamma$. (\textbf{b}) Upper panel: Periodic external bias as defined in Eq.\,\ref{eq:S_exBias} with $A=2, T=200$. Lower panel: Time-dependent response of the system as calculated by solving the set of differential equations with $\epsilon=0.5$. (\textbf{c}) Response of the system as a function of external bias. (\textbf{d}) Response of the system in frequency space, given by the Fourier transformation over a cycle of 100\,periods.
    \label{fig:S_Proxy_Schmitt}}
\end{figure}

We use the proxy system of \hyperref[Note_S:Dynamics_Instability]{Supplementary Note 2} to study the dynamics of the bistable OECT system. As before, we approach the Gibbs free energy function by the fourth order polynomial
\begin{equation}
    V(x)  = \alpha x^4 - \beta x^2 \quad \text{with}\quad \alpha = 0.25, \beta = 0.5.
\end{equation}
The dynamic response of this system is described by a set of two first-order differential equations,
\begin{align}
        \dot{x} &= v \\
    \dot{v} &= -\frac{\partial V(x)}{\partial x} - \epsilon v + \gamma,
\end{align}
with the damping coefficient $\epsilon$ and the external driving force $\gamma$. The latter causes a deflection of the potential function, as shown in Fig.\,\ref{fig:S_Proxy_Schmitt}a. The external driving force can be considered as time dependent, for instance by means of a periodic signal
\begin{equation}
\label{eq:S_exBias}
    \gamma(t) = A \cdot \sin\left(\frac{2\pi t}{T}\right) \quad \text{with} \quad A=2, T=200,
\end{equation}
shown in Fig.\,\ref{fig:S_Proxy_Schmitt}b (upper panel). Together with a damping coefficient of $\epsilon=0.5$, we solve the set of differential equations by numerical integration using the SciPy library\autocite{2020SciPy-NMeth}. Fig.\,\ref{fig:S_Proxy_Schmitt}b (lower panel) shows the time-dependent response of the system with corresponding damped oscillations towards the upper and lower state. This reflects even more in Fig.\,\ref{fig:S_Proxy_Schmitt}c, showing the response as a function of external bias. In the frequency domain, the higher harmonics are clearly visible, as to be expected for a system of intrinsic bistability (Fig.\,\ref{fig:S_Proxy_Schmitt}d). \\ \\
With this notion, we study the Fourier transformation of the experimental OECT data using a Fast-Fourier Transformation (FFT) algorithm. Therefore, the internal MATLAB function was used in the following form:
\\
\begin{lstlisting}[style=Matlab-editor]
% RecordedWaveform consists of time values (column 1) and current values (column 3)
L=length(RecordedWaveform(1:end,1));
fs=abs(1./(RecordedWaveform(2,1)-RecordedWaveform(1,1))); % sample rate
f4 = (fs)*(0:(L/2))/L;
SpeIn4=fft(RecordedWaveform(1:end,3));
P2in = abs(SpeIn4/L);
P1in = P2in(1:L/2+1);
P1in(2:end-1) = 2*P1in(2:end-1);
\end{lstlisting}

\begin{figure}[h!]
    \centering
    \includegraphics[width=0.98\linewidth]{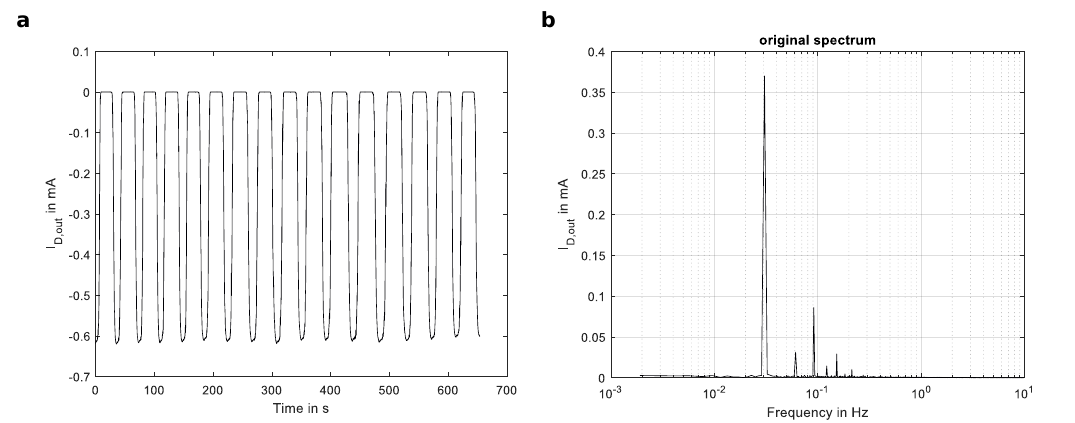} 
    \caption{\textbf{OECT oscillation.} (\textbf{a}) Recorded waveform of output current and (\textbf{b}) corresponding spectrum using an FFT algorithm in MATLAB.
    \label{fig:S_SchmittTrigger_1}}
\end{figure}

\noindent Note that the FFT assumes the signal in time domain with infinite iterations, i.e., the first value of the recorded waveform is seen as the follow-up of the last and vice versa. Hence, the recorded waveform including beginning and ending should result in an integer value to avoid spurs in the spectrum. Furthermore, a higher number of cycles should be recorded for a higher precision. A longer recording duration leads to a higher accuracy at lower frequencies, while a higher sample rate increases the accuracy at higher frequencies. We considered the above mentioned points for the calculation of the spectrum shown in Fig.\,\ref{fig:4}. Thus, we used a measurement consisting of 16\,periods (653 seconds recorded) with a sample rate of 19.6\,Hz as shown in Fig.\,\ref{fig:S_SchmittTrigger_1}a. The unfiltered spectrum is shown in Fig.\,\ref{fig:S_SchmittTrigger_1}b, showing higher harmonics, in line with our preliminary considerations of the proxy system.

\noindent For the filter design, we calculated several types of filters including higher order Butterworth and Elliptic filters. However, for an easy hardware realization, a simple filter structure ought to be used, e.g., a first-order Butterworth filter. This filter can be realized utilizing only two passive components, e.g., one resistor and one capacitor. For the design, we again used an internal MATLAB function as follows:
\\
\begin{lstlisting}[style=Matlab-editor]
fc4=6e-2; %corner frequency
[ym,xm] = max(P1in(2:end)); %skip dc
fc5 = f4(xm);
Wn4=2*fc4/fs;
[b4,a4] = butter(1,Wn4,'low'); %1st order Butterworth low-pass filter
nv4=filter(b4,a4,RecordedWaveform(1:end,3)); 
SpeOut4F=fft(nv4);
P2outF = abs(SpeOut4F/L);
P1outF = P2outF(1:L/2+1);
P1outF(2:end-1) = 2*P1outF(2:end-1);
[h_calc4, w_calc4] = freqz(b4,a4,L,fs); %filter's frequency response
\end{lstlisting}

\begin{figure}[h!]
    \centering
    \includegraphics[width=0.98\linewidth]{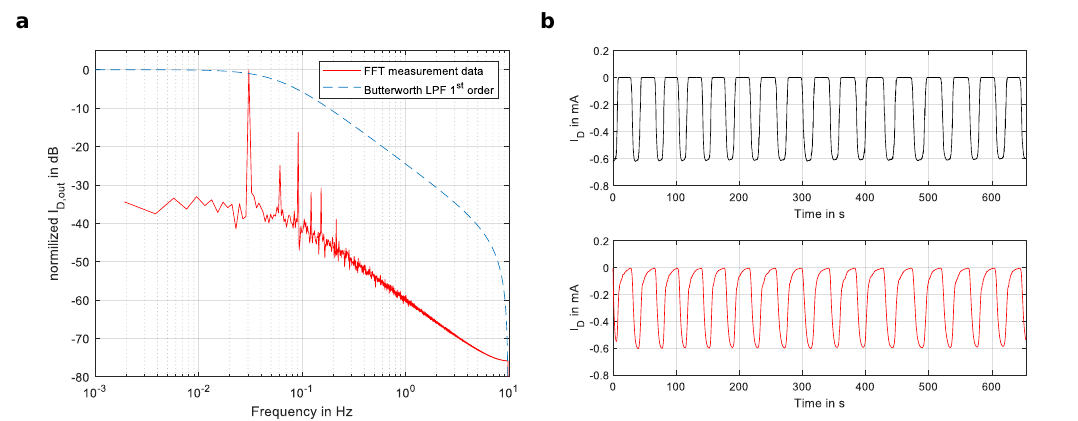} 
    \caption{\textbf{First-order Butterworth filter.} (\textbf{a}) Filtered spectrum in frequency domain using a first-order Butterworth filter (transfer function shown as dashed line) and recorded and filtered waveforms of the output current in time domain (\textbf{b}).
    \label{fig:S_SchmittTrigger_2}}
\end{figure}

\noindent This results in a filter function with a 3\,dB-corner frequency of 60\,mHz, a passband gain of 0\,dB (passive filter), and a stop band attenuation of 20\,dBdec$^{-1}$  as shown in Fig.\,\ref{fig:S_SchmittTrigger_2}a. There, the filtered spectrum is also shown in frequency domain, while the filtered and unfiltered signals in time domain are shown in Fig.\,\ref{fig:S_SchmittTrigger_2}b.

\noindent The obtained filter coefficient can be implemented, e.g., by a $24\,\si{\kilo\Omega}$ resistor and a $1.1\si{\milli\farad}$ capacitor, both of which are commercially available.

\clearpage

\section*{Supplementary References}
\printbibliography[heading = none]